\documentclass[
showpacs,
amsmath,amssymb,
aps,
prb,twocolumn,
floatfix,
nofootinbib
]{revtex4-2}

\usepackage{graphicx}% Include figure files
\usepackage{amsmath}
\usepackage{dcolumn}% Align table columns on decimal point
\usepackage{bm}% bold math
\usepackage{xcolor}
\usepackage{ulem} 

\newcommand{\Z}{\mathbb{Z}}

\begin{document}

\title{Kapitsa pendulum effects in Josephson junction + nanomagnet under external periodic drive}
%\title{Kapitsa pendulum effects in nanomagnet-Josephson junction system}
%\title {Bessel dependence of the reorientation voltage in nanomagnet-Josephson junction system}

\author{K. V. Kulikov$^{1,2}$, D. V. Anghel$^{1,3}$, A. T. Preda$^{3,4}$, M. Nashaat$^{1,5}$, M. Sameh$^{5}$ and Yu. M. Shukrinov$^{1,2,6}$ }

\affiliation{$^1$ BLTP, JINR, Dubna, Moscow region, 141980, Russia\\
	$^2$ Dubna State University, Dubna, Russia\\
	$^3$ \mbox{Horia Hulubei National Institute for R\& D in Physics and Nuclear Engineering, M\u{a}gurele, Romania}\\
	$^4$ \mbox{University of Bucharest, Faculty of Physics, Bucharest, Romania} \\
	$^5$ \mbox{Department of Physics, Faculty of Science, Cairo University, 12613, Giza, Egypt}\\
	$^6$ \mbox{Moscow Institute of Physics and Technology, Dolgoprudny, 141700 Russia}
	}
\date{\today}

\begin{abstract}
We investigate reorientation effects under external periodic drive in the nanomagnet dynamics coupled to a Josephson junction. The Kapitsa pendulum is introduced as a mechanical analog to this system and we demonstrate the reorientation of the easy axis of the nanomagnet. The magnetic field generated by the Josephson junction and external drive plays the role of the oscillating force of the suspension point in the Kapitsa pendulum. The high frequency oscillations change the orientation of the magnetic moment. The magnetic field of the quasiparticle current determines the frequency dependence of the magnetic moment's orientation. We obtain simple analytical formulas for the stable position of the magnetic moment, both under the external periodic drive and without it. The influence of external periodic drive on the voltage of complete reorientation have been demonstrated.
\end{abstract}


\maketitle

%\paragraph*{Introduction.}
\section{Introduction} \label{sec_intro}

Kapitsa's pioneering work ~\cite{kapitza} initiated the field of vibrational mechanics, and his method is used to describe periodic processes in a variety of different physical systems, like atomic physics~\cite{bukov15, borromeo07, aidelsburger14, Wickenbrock2012},  plasma physics,   optics \cite{Chizhevsky2014}, condensed matter physics,  biophysics \cite{Uzuntarla2015} and cybernetical physics (see [\onlinecite{boukobza2010,citro2015,fialko15,longhi17,shayak17,Martin18}] and references therein). In particular, imposing vibrational quantum coherence into topological states of matter may become a universal light control principle for reinforcing the symmetry-protected helical transport \cite{yang20}. Coherent lattice vibrations can have direct and profound effects on surface transport of Dirac fermions, via periodic modulation of electronic states. Ultrafast phononics has been explored as a new avenue to manipulate properties of superconductors, oxides, semimetal and photovoltaic semiconductors. In a chain of spins with long-range ferromagnetic interactions, a magnetic field with periodic modulation can lead to stability regions of ferromagnetic spins around unstable paramagnetic configuration \cite{lerose_prb-19}.  In nonlinear control theory, the Kapitsa pendulum is used as an example of a parametric oscillator that demonstrates the concept of ``dynamic stabilization''.

In Ref.~[\onlinecite{richards_sr-18}], the authors realized experimentally the Kapitsa pendulum at the micrometre scale using a colloidal particle suspended in water and trapped by optical tweezers.
Moreover, it was analytically and experimentally demonstrated that if the oscillation direction of the pendulum suspension point change over time, so does the pendulum equilibrium point and active damping control can take place.
The Kapitsa quantum pendulum can be stabilized in the form of quantum states near a local minimum of the effective potential energy \cite{Golovinski2021}.  

The coupling mechanisms of Josephson junctions (JJs) to magnets (or individual spins) in proximity of each other have been intensively studied in the past.  The theory traces back to the year 1966 to the works of Kulik \cite{kulik1966zh} and Bulaevskii et al.~\cite{bulaevskii1977pis}, where they clarified the effect of magnetic moment flips on the tunneling current. Since then, a number of works were published describing the interaction of spin with superconducting correlations inside a Josephson junction \cite{zhu2004novel, nussinov2005spin},  the Josephson current through a multilevel quantum dot with spin-orbit coupling \cite{dell2007josephson}, the formation of vortices in a Josephson junction by a magnetic dot \cite{samokhvalov2009current} and spin-orbit coupling of a single spin to the Josephson junction \cite{padurariu2010theoretical}, etc. 

The peculiarity of the presented system is manifested in the choice of the geometry of the structure, the nature of the interaction and the resistance in the normal state, which is taken into account in the framework of the resistively shunted Josephson junction model (RSJ-model)  \cite{likharev1986dynamics}. The attractiveness of the model with purely electromagnetic interaction lies in the absence of unknown parameters, which should be essential for its experimental study. Within the framework of this model, it is assumed that a number of characteristic phenomena will be observed, in particular, the appearance on the I-V characteristic of JJs of Shapiro-like steps created by the NM precession \cite{cc-prb_10, ghosh2017magnetization}; the reversal of the magnetic moment when a voltage or a current that changes in time is applied to the JJ \cite{cc-prb_10, snrk-jetpl_19}; and also the Rabi oscillations of the quantum spin induced by the applied constant voltage. A remarkable property of the system is that, despite the weakness of the field generated by the tunneling current, at a certain time dependence of the applied voltage, an effective pumping of spin excitations into the nanomagnet can be realized as well as a reversal of the magnetic moment. Rabi oscillations of the quantum spin in the JJ-NM system, induced by an applied constant voltage, are determined by the ratio of the Zeeman interaction of the spin with the tunneling current field and the tunneling splitting $ \Delta $ \cite{cc-prb_10}. They are highly dependent on the applied voltage $V_0$. The greatest effect occurs when $ V_0 $ satisfies one of the resonance conditions $ eV_0 = (m/n) \Delta $, where $ m $ and $ n $ are integers. With such a resonant behavior, the probability of finding a spin in the up or down state is very different from the nonresonant case, which indicates the fundamental possibility of electromagnetic control of the JJ-NM qubit by means of an applied voltage.

Early experimental research on coupling of magnetic nanoparticles to SQUIDs has been reviewed by Wernsdorfer \cite{wernsdorfer2006classical}. 
The possibility of switching the magnetization of Co nanoparticles in a dc magnetic field by the rf pulse has been demonstrated by Thirion et al. \cite{thirion2nature}.
Further miniaturization of such systems has been achieved using carbon nanotubes \cite{cleuziou2006carbon, cleuziou2007gate, bogani2010effect} and nanolithography assisted by the atomic force microscope \cite{faucher2009optimizing}. 
These systems utilizing single nanomagnets which possess very different magnetic properties from bulk material may provide advanced replacements for hard disk media \cite{wu1998large} and computer memory chips \cite{prinz1998magnetoelectronics}. 

In Ref.~[\onlinecite{snrk-jetpl_19}], the authors introduced the Kapitsa pendulum as a mechanical analog to the JJ-nanomagnet system and demonstrated the reorientation of the easy axis of the magnetic moment of the nanomagnet. In this case, the Josephson to magnetic energy ratio $G$ corresponds to the amplitude of the variable force of the Kapitsa pendulum, the Josephson frequency $\Omega_J$ corresponds to the oscillation frequency of the suspension point, and the averaged magnetic moment components specify the stable position. Howewer, the results showed that the reorientation value was in reverse proportion to the frequency of the force applied to the suspension point. In~[\onlinecite{snrk-jetpl_19}], the increase in $\Omega_J$ leads to a larger reorientation value at the same value of $G$. On the other hand, an opposite result has been observed in Ref.~[\onlinecite{smrbb-epl_18}], where for the $\varphi_{0}$-junction the decrease in $\Omega_J$ led to the larger reorientation value.

In this paper we study the dynamics of a nanomagnet coupled to a Josephson junction (see Fig.\ref{1}(a)) under external periodic drive. We show an important role of  the quasiparticle current in the effective field of the system: it can change the frequency dependence of the Kapitsa pendulum characteristics. We also investigate the effect of external drive on the reorientation of the nanomagnet easy axis.

%\paragraph*{Model and Methods.}
\section{Model and Methods} \label{sec_MM}

\begin{figure}[t]
		\includegraphics[width=0.52\linewidth]{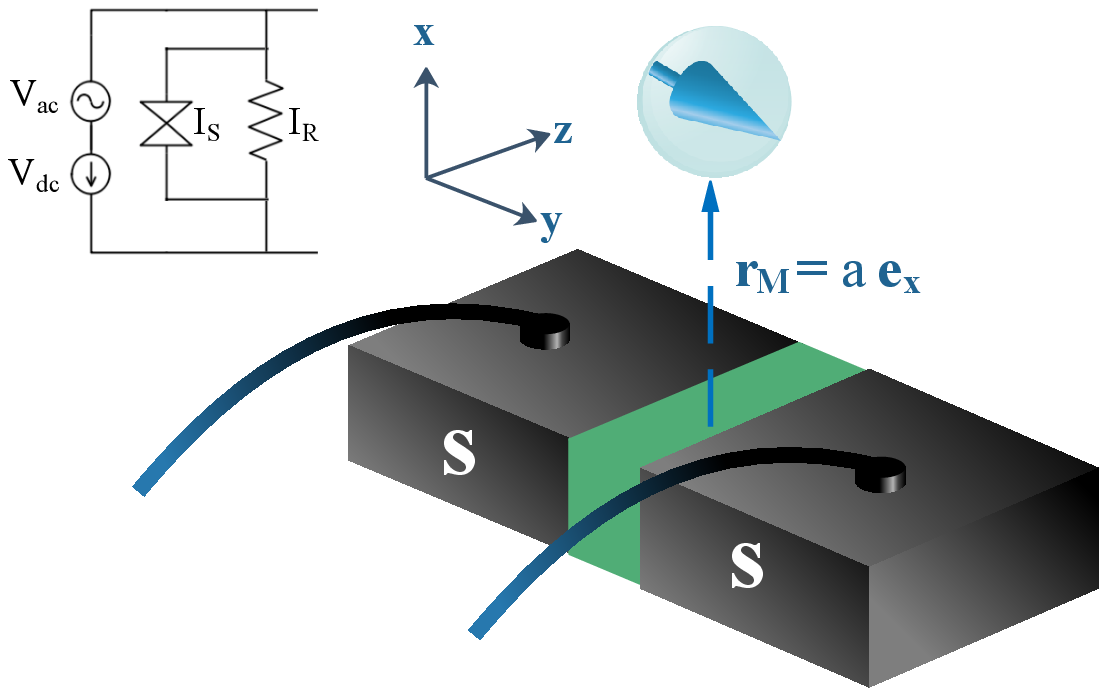}\includegraphics[width=0.38\linewidth]{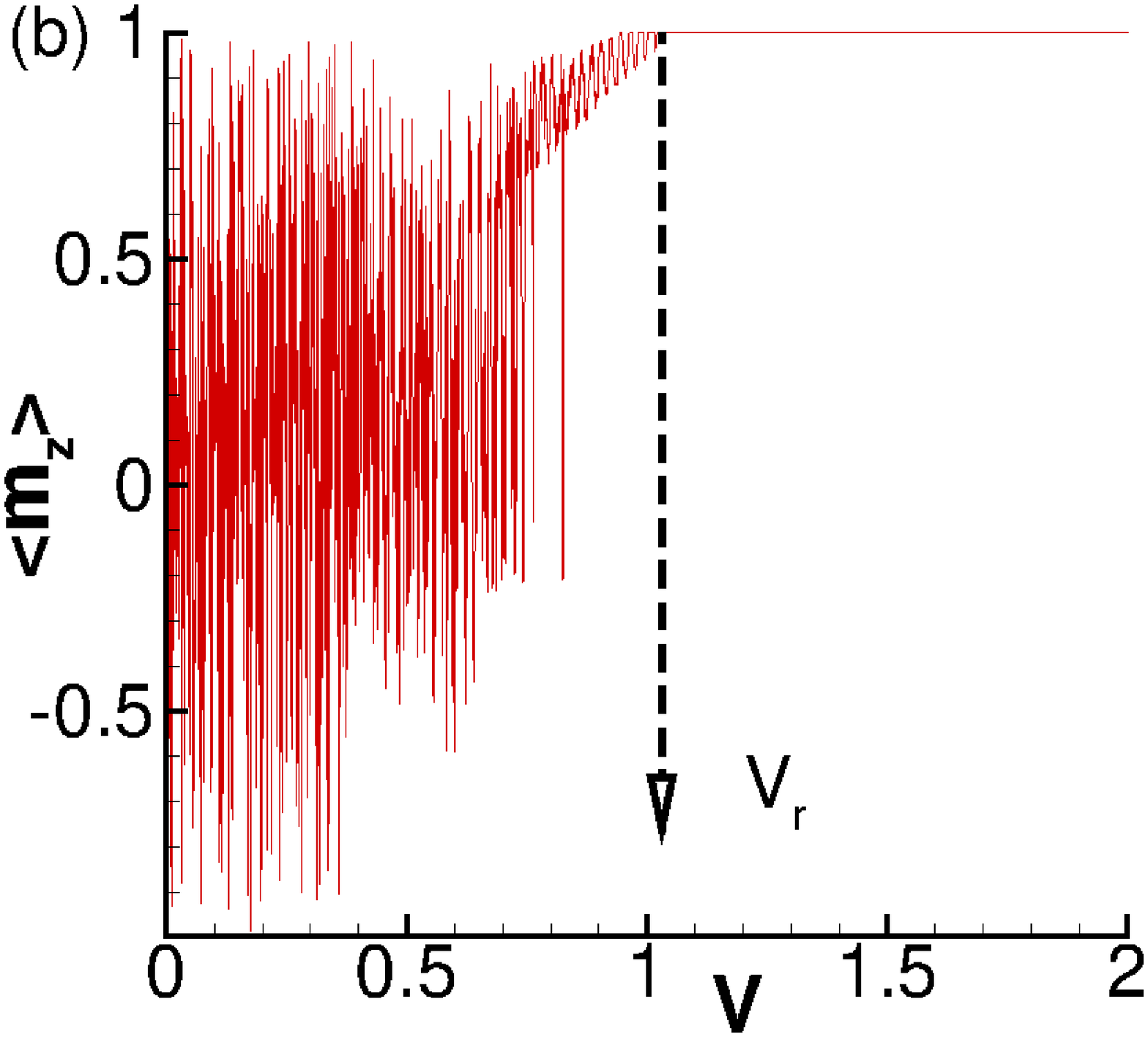}
		\caption{(a) Schematic diagram of the considered system with the system geometry. (b) Average  magnetic moment component $m_z$ of the nanomagnet as a function of the voltage across the JJ demonstrating magnetization reorientation in voltage bias regime at $\alpha = 0.1$, $G = 3\pi$, $k = 0.1$, $\Omega_F = 0.5$ and $\Omega = 0.8$. The dashed arrow indicates the voltage of complete reorientation.  }
		\label{1}
\end{figure}

We consider a voltage biased short Josephson junction of length $l$ coupled to a nanomagnet with magnetic moment $\textbf{M}=(M_x,M_y,M_z)$ located at distance $\textbf{r}_{M}=a \textbf{e}_x$ from the center of the junction as shown in Fig.\ref{1}(a). The interaction between the two systems is considered to be of purely electromagnetic origin. The magnetic field of the nanomagnet alters the Josephson current flowing through the junction while the magnetic flux generated by the Josephson junction acts on the magnetic moment of the nanomagnet. This model was developed in Ref. \cite{cc-prb_10}.

The dynamics of magnetic moment  can be described by Landau-Lifshitz-Gilbert (LLG) equation \cite{lifshitz1981course}
\begin{eqnarray}
\dfrac{d\textbf{M}}{d\tau} = \gamma \textbf{H}_{eff} \times \textbf{M}+\dfrac{\alpha}{M_s}\left(\textbf{M}\times\dfrac{d\textbf{M}}{d\tau}\right),
\label{LLG_components}
\end{eqnarray}
where $\alpha$ is the Gilbert damping parameter, $\textbf{H}_{eff}$ is the effective field with components $(H_x, H_y, H_z)$ and $\gamma$ is the gyromagnetic ratio.

The effective field acting on the nanomagnet is given by \cite{lifshitz1981course} 

\begin{eqnarray}
\textbf{H}_{eff}=-\dfrac{1}{v}\dfrac{\partial E}{\partial\textbf{M}},
\label{Effective_Field_1}
\end{eqnarray}
where $E=E_M+E_Z$ \cite{likharev1986dynamics} is the total energy of the system, where
\begin{eqnarray}
E_M=\frac{- K v}{2} (M_{y}/M_{s})^{2},
\label{Effective_Field_M}
\end{eqnarray}
is the anisotropy energy of the nanomagnet, $K$ is the anisotropy constant, $v$ is the volume of nanomagnet, $M_{y}$ is the magnetization in y-direction (easy axis), $M_{s}$ is the value of saturation magnetization,
\begin{equation}
E_Z=E_{Z,n}+E_{Z,s}=-I \int d\textbf{r} \textbf{A}_M(\textbf{r},\tau),
\label{Effective_Field_Z}
\end{equation}
is the Zeeman energy, $I$ is the total current, $E_{Z,n}$ and $E_{Z,s}$ are the Zeeman terms related to normal and superconducting currents, respectively, 
\begin{equation}
\textbf{A}_M(\textbf{r},\tau)=\dfrac{\mu_0}{4\pi}\dfrac{\textbf{M}(\tau)\times\textbf{r}}{r^3}
\label{vector_poten}
\end{equation}
is the vector potential created at position $r$ from the nanomagnet, which is assumed to be much smaller than all other dimensions of the problem, $\mu_0$ is the vacuum permeability,
\begin{equation}
E_{Z,n}=-I_N \int d\textbf{r} \textbf{A}_M(\textbf{r},\tau),
\label{Effective_Field_Z_n}
\end{equation}
where $I_N$ is the quasiparticle current,
\begin{eqnarray}
E_{Z,s}=E_J=-\epsilon_J \cos(\varphi),
\label{Effective_Field_J}
\end{eqnarray}
is the energy of the JJ, $\epsilon_J=\hbar I_c/(2e)$, $I_{c}$ is the critical current of the JJ and $\varphi$ is the phase difference of the JJ.

In our model we consider the short JJ, in which the width $L$ is less than the Josephson penetration depth $\lambda_J$. So, the gradient of the phase difference along the barrier is absent. As we mention earlier, the magnetic field, created by the rotating magnetic moment of the nanomagnet alters the current flowing through the junction. Therefore, a shift of the phase difference in the JJ arises. For the JJ biased by the ac and dc voltage the total phase difference is given by 
\begin{equation}
\varphi = \varphi_0 + \varphi_a,
\label{total_phase}
\end{equation}
where
\begin{equation}
\frac{d\varphi_0}{d\tau} = \frac{2e}{\hbar}(V_{dc}+V_{ac}\cos(\omega \tau))
\label{phase_dif}
\end{equation}
and
\begin{equation}
\varphi_a = - \dfrac{2\pi}{\Phi_0} \int d\textbf{r} \textbf{A}_M(\textbf{r},\tau)
\label{phase_shift}
\end{equation}
is the phase difference induced at the junction by the time-dependent magnetic field generated by the rotating magnetic moment. The integration is carried out along the 1D current path in the JJ (from $0$ to $l$), so $dr$ is an element of this path. Here $\Phi_{0}$ is the flux quantum, $V_{dc}$ is the dc voltage bias, $V_{ac}$ is the ac voltage bias and $\omega$ is the frequency of the external drive. Substituting (\ref{vector_poten}) into (\ref{phase_shift}) one can get

\begin{eqnarray}
\varphi_a &=& - \dfrac{\mu_0}{2\Phi_0} \int d\textbf{r} \dfrac{\textbf{M}\times\textbf{r}}{r^3} \nonumber\\ 
&=& - \dfrac{\mu_0M_{s}}{2\Phi_0} \dfrac{l}{a\sqrt{l^2+a^2}} m_{z} =-km_z,
\label{Phase_dif_nm}
\end{eqnarray}
where $k$ is the coupling constant between the JJ and the nanomagnet, whereas $m_z$ is magnetization component normalized to $M_s$.

The total current through the JJ in the RSJ-model in the dimensionless form is given by

\begin{eqnarray}
I &= \sin(V t + \varphi_a+\dfrac{A}{\Omega}\sin(\Omega t))\nonumber\\ &+  V+A \cos(\Omega t) +   \dot \varphi_a,
\label{Total_current}
\end{eqnarray}
where $I$ is normalized to $I_c$, $t=\tau\omega_c$, $V$ is the dc voltage bias normalized to $V_c=\hbar \omega_c/ 2e$, $\omega_{c}$=$2eRI_{c}/\hbar$ is characteristic Josephson frequency, $R$ is the resistance of the JJ in normal state, $A=V_{ac}/V_c$ is the amplitude of external drive and $\Omega$ is the frequency of the external drive normalized to $\omega_c$.

By substituting (\ref{Total_current}) and (\ref{Phase_dif_nm}) into (\ref{Effective_Field_Z}), then using the Biot-Savart law to calculate the magnetic field acting on the nanomagnet generated by the Josephson junction, one can find components of the effective field as

\begin{eqnarray}
h_{x}&=&0, h_{y} = m_{y},  \nonumber\\ h_{z}& =& \epsilon [\sin(V t - k m_{z}+\frac{A}{\Omega}\sin(\Omega t))\nonumber\\&+& \delta( V+A \cos(\Omega t) - k \dot m_{z})].
\label{Effective_Field_comp}
\end{eqnarray}
Here we use normalized units, $m_{i}=M_{i}/M_{s}$ ($i=1,2,3$), $h_{i}=H_{i}/H_{F}$, where $H_{F}=\Omega_{F}/\gamma$,  $\Omega_{F}=\omega_F/\omega_c$ is frequency of the ferromagnetic resonance, $\epsilon=Gk$,  $G=\epsilon_J / K v$ is the Josephson to magnetic energy ratio, whereas $\delta = 1, 0$ is the parameter which we use to indicate the terms that come from quasiparticle current. The case with $\delta=0$ is studied in Ref.~\cite{smrbb-epl_18}, where the authors demonstrate the reorientation of the easy axis in $\varphi_{0}$-junction with inverse Kapitsa-like pendulum feature. Here, we consider $\delta$ = 1, in which we take into account the effect of the superconducting and quasiparticle tunneling currents. Notice also that in our normalization, the dc voltage bias $V$ is equal to the Josephson frequency $\Omega_J$.

We use Eq. (\ref{LLG_components}) with the effective field (\ref{Effective_Field_comp}) to numerically calculate the dynamics of the magnetic moment projections on the coordinate axes, with the initial conditions $m_{x}=0$, $m_{y}=1$, and $m_{z}=0$.

As in a Kapitsa pendulum \cite{landau-mech,kapitza}, the applied voltage across the Josephson junction in our system generates a high frequency magnetic field that reorients the magnetic moment of the nanomagnet. Figure \ref{1}b  shows the reorientation of the magnetic moment as a function of the dc bias voltage, i.e., a manifestation of the Kapitsa pendulum feature in the Josephson junction--nanomagnet system. The stabilization of the magnetic moment components dynamics occurs at $M = (0,0,1)$, when $V$ exceeds a certain reorientation value $V_r$. This represents a complete reorientation of the magnetic moment.

%\paragraph*{Analytical description}
\section{Analytical description} \label{sec_analytic}

We study analytically the magnetic moment dynamics of the nanomagnet in the approximation $V, \Omega \gg \Omega_F$, that is, when the frequencies of the JJ and of the external drive are much higher than the eigen-frequency of the nanomagnet.
This produce small and fast oscillations of the magnetic moment, similar to the oscillations in the Kapitsa pendulum \cite{landau-mech}. The calculations details are given in the Appendix. We use the spherical coordinates $\theta, \phi$ to write $(m_x,m_y,m_z) \equiv (\sin \theta \cos \varphi,\sin \theta \sin \varphi, \cos \theta )$ and separate them into fast and slow variables by introducing the notations $\theta \equiv \Theta + \xi$ and $\phi \equiv \Phi + \zeta $. Here, $\Theta$ and $\Phi$  describe the ``slow'' motion, relevant on longer time scales (comparable to the period of the oscillations of the system in the absence of the external drive and the Josephson oscillations), whereas the variables $\xi$ and $\zeta$ describe the ``fast'' oscillations of the system, which take place on shorter time scales (comparable to $1/V$ and $1/\Omega$).
Writing Eq.~(\ref{LLG_components}) in the variables $(\theta,\phi)$ (see Appendix~\ref{sec_BE}) and expanding it in a Taylor series to the first order in $(\xi,\zeta)$ around $(0,0)$ (see Appendix~\ref{sec_app_K}, Eqs.~\ref{dot_xi_zeta}), we obtain
\begin{subequations} \label{dot_xi_zeta_main}
	\begin{eqnarray}
	&& \dot\theta = \dot\Theta + \dot\xi
	\approx M_0(\Theta) F(\Theta, \Phi, \xi, \zeta, t)
	%   \nonumber \\
	%   %
	%   &\equiv& A_0(\Theta, \Phi) + A_1(\Theta, \Phi) \xi + A_2(\Theta, \Phi) \zeta + A_3(\Theta, \Phi) \tilde g_z(\Theta, t)
	\label{dot_xi_main}\\
	&& \dot\phi = \dot\Phi + \dot\zeta
	\approx M_0(\Theta) Q(\Theta, \Phi, \xi, \zeta, t)
	\label{dot_zeta_main}
	\end{eqnarray}
\end{subequations}
with $F(\Theta, \Phi, \xi, \zeta, t)= F_0(\Theta, \Phi) + F_{\xi}(\Theta, \Phi, t) \xi + F_{\zeta}(\Theta, \Phi, t) \zeta + F_t(\Theta,t)$ and $Q(\Theta, \Phi, \xi, \zeta, t)=Q_0(\Theta, \Phi) + Q_{\xi}(\Theta, \Phi, t) \xi + Q_{\zeta}(\Theta, \Phi, t) \zeta + Q_t(\Theta,t)$, where
\begin{widetext}
\begin{subequations} \label{F_dot_theta}
	\begin{eqnarray}
	&&M_0(\Theta) \equiv \frac{ \Omega_F }{ 1 + \alpha^2 + \delta \alpha \epsilon k \sin^2\Theta \, \Omega_F }, \quad
	F_0(\Theta, \Phi) = ( \alpha \sin\Phi \cos\Theta + \cos\Phi) \sin\Theta \sin\Phi - \alpha \epsilon \delta V \sin\Theta ,
	\label{F_theta} \\
	&&F_{\xi}(\Theta, \Phi, t) = \Big[- \cos\Phi \cos\Theta \Big(\sin^2\Theta  \alpha \delta \epsilon k \Omega_F  - \alpha^2 - 1 \Big) + 2 \alpha ( \alpha^2 + 1) \sin\Phi \cos^2\Theta  - \alpha (\alpha \delta \epsilon k \Omega_F \sin^2\Theta + \alpha^2 + 1) \sin\Phi \Big] \nonumber\\
	&& \times\frac{\sin\Phi}{ 1 + \alpha^2 + \delta \alpha \epsilon k \sin^2\Theta \, \Omega_F }- \alpha \epsilon k \sin^2\Theta \cos\left[ V t - k \cos\Theta + \frac{A}{\Omega} \sin(\Omega t) \right] ,
	\label{F_theta_xi} \\
	&&F_{\zeta}(\Theta, \Phi, t) = (2 \cos\Phi  \sin\Phi  \cos\Theta  \alpha + 2 \cos^2\Phi  - 1) \sin\Theta ,
	\label{F_theta_zeta} \\
	&&F_t(\Theta,t) = - \alpha \sin\Theta \epsilon \delta A \cos(\Omega t)- \alpha \epsilon \sin\Theta  \sin\left[ V t - k \cos\Theta + \frac{A}{\Omega} \sin(\Omega t) \right] , \label{F_theta_time} \\
	&&Q_0(\Theta, \Phi)  = \Big[\epsilon \delta V + (\delta \epsilon k \Omega_F \sin^2\Theta \cos\Phi- \cos\Theta \sin\Phi + \alpha \cos\Phi) \sin\Phi \Big] ,
	\label{Q_0} \\
	&&Q_{\xi}(\Theta, \Phi, t)=\left\{\left(\frac{\left[\Omega_{F} \alpha \delta \epsilon k\left(\cos ^{2} \Theta+1\right)+\alpha^{2}+1\right] \sin \Phi+2 \cos \Phi \cos \Theta \delta \epsilon k \Omega_{F}}{1+\alpha^{2}+\delta \alpha \epsilon k \sin ^{2} \Theta \Omega_{F}}\right) \sin \Phi\right. \nonumber \\
	&&\left.+\epsilon k \cos \left[V t-k \cos \Theta+\frac{A}{\Omega} \sin (\Omega t)\right]\right\} \sin \Theta ,
	\label{Q_phi_xi} \\
	&&Q_{\zeta}(\Theta, \Phi, t)  = \Big[ (2\cos^2\Phi  - 1) (k \delta \epsilon \Omega_F \sin^2\Theta + \alpha)- 2 \cos\Phi  \cos\Theta  \sin\Phi \Big] ,
	\label{Q_phi_zeta} \\
	&&Q_t(\Theta, t)  = \epsilon \delta A \cos(\Omega t)+ \epsilon \sin\left[ V t - k \cos\Theta + \frac{A}{\Omega} \sin(\Omega t) \right] ,
	\label{Q_phi_time}
	\end{eqnarray}
\end{subequations}
with (Eqs.~\ref{exp_sin_cos_Bessel2_main})
\begin{subequations} \label{exp_sin_cos_Bessel2_main}
\begin{eqnarray}
	\sin[ V t - k \cos\Theta + \frac{A}{\Omega} \sin( \Omega t )] &=& \sum_{m=-\infty}^{\infty} \text{sign}^m(m) J_{|m|}\left(\frac{A}{\Omega}\right)
	\sin[ (V + m \Omega) t - k \cos\Theta ] ,
	\label{def_g1pg2_main} \\
	\cos[ V t - k \cos\Theta + \frac{A}{\Omega} \sin( \Omega t ) ] &=& \sum_{m=-\infty}^{\infty} \text{sign}^m(m) J_{|m|}\left(\frac{A}{\Omega}\right)
	\cos [ ( V + m \Omega ) t - k \cos\Theta ] .
	\label{def_h1ph2_main}
\end{eqnarray}
\end{subequations}
\end{widetext}
In the definitions (\ref{F_dot_theta}), $M_0$, $F_0$, and $Q_0$ do not depend explicitly on time.
Plugging Eqs.~(\ref{exp_sin_cos_Bessel2_main}) into (\ref{F_dot_theta}), we see that, if $A>0$, $F_t$ and $Q_t$ may be written as infinite sums of terms that oscillate with the frequencies $\Omega$ or $V+m\Omega$.
We denote $m_0\equiv-V/\Omega$.
If $m_0$ is an integer, we have a \textit{zeroth order resonance} (this case was discussed in detail in Appendix~\ref{subsubsec_stability_m0}).
Then, we introduce the notation
%
%\begin{subequations} \label{defs_Qm0_Fm0}
%\begin{eqnarray}
%	Q_{m_0}(\Theta) &\equiv&
%	-\epsilon \text{sign}^{m_0}(m_0) J_{|m_0|}\left(\frac{A}{\Omega}\right) \sin[k\cos\Theta] , \label{defs_Qm0} \\
%	%
%	F_{m_0}(\Theta) &\equiv& - \alpha \sin\Theta \, Q_{m_0}(\Theta) , \label{defs_Fm0}
%\end{eqnarray}
%\end{subequations}
\begin{equation}
	Q_{m_0}(\Theta) \equiv -\epsilon \text{sign}^{m_0}(m_0) J_{|m_0|}\left(\frac{A}{\Omega}\right) \sin[k\cos\Theta] , \label{defs_Qm0_Fm0}
\end{equation}
and we may write the slow motion velocity in the zeroth order of approximation (~\ref{dot_Theta1_Phi1}, \ref{dot_Theta_Phi_0})

\begin{equation}
	\begin{aligned}
	& \dot{\Theta}_0 \equiv M_0(\Theta) \left[ F_0(\Theta,\Phi) - \alpha \sin\Theta \, Q_{m_0}(\Theta) \right] , \\
	& \dot{\Phi}_0 \equiv M_0(\Theta) \left[ Q_0(\Theta,\Phi) + Q_{m_0}(\Theta) \right] ,
	\end{aligned}
	\label{dTdP0}
\end{equation}
where $Q_{m_0}(\Theta)=0$ if $m_0 \notin \Z$ or $A=0$.
Since $\Omega,V \gg \Omega_{F}$, and if $|V+m\Omega|\gg\Omega_{F}$ for any $m\ne m_0$, the terms that depend explicitly on time in $F_t$ and $Q_t$ (~\ref{F_dot_theta}) oscillate fast as compared to the period of oscillation in the absence of perturbation and their contribution to $(\Theta, \Phi)$ averages to zero.
So, we have in general (Appendix~\ref{sec_app_K})
%
%\begin{eqnarray}
%	&& \dot{\zeta}_0(\Theta,t) \equiv M_0(\Theta) [Q_t(\Theta,t) - Q_{m_0}(\Theta)] ,\nonumber \\
%	&& \dot{\xi}_0(\Theta,t) = -\alpha\sin\Theta\, \dot{\zeta_0},
%	\nonumber \\
%	%
%	&& \zeta_0(\Theta,t) = \int_{0}^{t}\dot{\zeta}_0(\Theta,t) \, dt , \ \xi_0(\Theta,t) = \int_{0}^{t}\dot{\xi}_0(\Theta,t) dt.
%	\label{dxidz0}
%\end{eqnarray}
\begin{equation}
	\begin{aligned}
		& \dot{\zeta}_0(\Theta,t) \equiv M_0(\Theta) [Q_t(\Theta,t) - Q_{m_0}(\Theta)] , \\
		& \dot{\xi}_0(\Theta,t) = -\alpha\sin\Theta\, \dot{\zeta_0}, \\
		& \zeta_0(\Theta,t) = \int_{0}^{t}\dot{\zeta}_0(\Theta,t) \, dt , \ \xi_0(\Theta,t) = \int_{0}^{t}\dot{\xi}_0(\Theta,t) dt.
	\end{aligned}
	\label{dxidz0}
\end{equation}

In the order $n\ge 1$ of approximation we have
%
%\begin{subequations} \label{dxidz1}
%\begin{equation}
%	\dot{\xi}_n = \dot{\xi}_{n-1} + M_0(\Theta) [F_{\xi}(\Theta, \Phi, t) \xi_{n-1} + F_{\zeta}(\Theta, \Phi, t) \zeta_{n-1}],
%	\label{dxi1}
%\end{equation}
%%
%\begin{equation}
%	\dot{\zeta}_n = \dot{\zeta}_{n-1} + M_0(\Theta) [Q_{\xi}(\Theta, \Phi, t) \xi_{n-1} + Q_{\zeta}(\Theta, \Phi, t) \zeta_{n-1}] . \label{dz1}
%\end{equation}
%\end{subequations}

\begin{equation}
	\begin{aligned}
	& \dot{\xi}_n = \dot{\xi}_{n-1} + M_0(\Theta) [F_{\xi}(\Theta, \Phi, t) \xi_{n-1} + F_{\zeta}(\Theta, \Phi, t) \zeta_{n-1}], \\
	& \dot{\zeta}_n = \dot{\zeta}_{n-1} + M_0(\Theta) [Q_{\xi}(\Theta, \Phi, t) \xi_{n-1} + Q_{\zeta}(\Theta, \Phi, t) \zeta_{n-1}] .
	\end{aligned}
	\label{dxidz1}
\end{equation}

If $(\dot{\xi}_{n-1}, \dot{\zeta}_{n-1})$ contain terms oscillating with frequencies $nV+m\Omega$, then, from (\ref{F_dot_theta}), (\ref{dxidz1}), and (\ref{exp_sin_cos_Bessel2_main}) we observe that
$(\dot{\xi}_{n},\dot{\zeta}_{n})$ contain terms oscillating with frequencies $(n+1)V+m\Omega$, for any $m$, such that  $(n+1)V+m\Omega\ne 0$.
If exists an integer $m^{(n)}_{0}$, such that $(n+1)V+m^{(n)}_0\Omega = 0$, this would give a term which does not explicitly depend on time and therefore is incorporated into the $n^{\rm th}$ order contribution to the slow motion $(\dot{\Theta}_n, \dot{\Phi}_n)$.
This contribution contain a factor $M_0^n(\Theta)$, as compared to $(\dot{\Theta}_0, \dot{\Phi}_0)$, which decreases to zero with $n$ if $M_0(\Theta)<1$ (we have, in  general, $\Omega_{F}\ll 1$).

\begin{figure}[t]
	\centering
	\includegraphics[width=6cm]{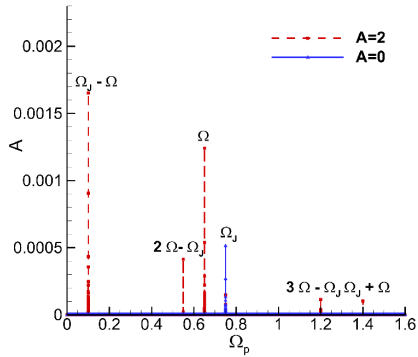}
	\caption{(Color online) Results of FFT analysis of $m_{z}(t)$ at $V=0.75$ without and with external drive of amplitude $A=2$ and frequency $\Omega=0.65$.}
	\label{fig:case1}
\end{figure}

The presence of the terms that oscillate in time may be emphasized by the fast Fourier transform (FFT) of $m_{z}(t)$.
The oscillating frequencies should be $|nV+m\Omega|$, where $n, m$ are integers and $n \ge 0$.
This can be clearly seen in Fig.~\ref{fig:case1}, where we present the results of $m_{z}(t)$'s precession frequency $\Omega_{p}$  for $V=0.75$ and $\Omega=0.65$, the other parameters being $A=0$ or $2$, $\Omega_{F}=10^{-2}$, $G=2\pi$, $k=0.1$ and $\alpha=0.1$.

Due to the coupling between JJ and nonomagnet through the current phase relation, at $A=0$, the magnetization $m_{z}$ oscillates only with the Josephson frequency $V$, hence we see only one frequency line (blue line) at $\Omega_{p}=V$.

If $A\neq0$ we have two frequencies, $V$ and $\Omega$, which affect the precession of $m_{z}(t)$.
In this case, we observe two harmonic frequency lines at $\Omega_{p}=V$ and  $\Omega_{p}=\Omega$. In addition, the FFT analysis gives several frequency lines corresponding to the subharmonics with $\Omega_{p} =n V \pm m \Omega$,  where $n$ and $m$ are integers.

%\paragraph*{Stability position in the case without periodic drive.}
\section{Stability position in the case without periodic drive} \label{sec_stability}

When $A = 0$, ~(\ref{F_dot_theta}) and (\ref{exp_sin_cos_Bessel2_main}) simplify considerably (see Section~\ref{subsubsec_A0}) and the equilibrium condition $\dot\Theta = \dot\Phi = 0$ give (see Appendix~\ref{subsubsec_A0})
\begin{subequations} \label{Eq_cond_A0_main}
	\begin{equation}
	\Phi = \pi/2 \qquad {\rm or} \qquad \Phi = 3\pi/2  \label{Eq_cond_A0_Phi_main}
	\end{equation}
and an equation for $\Theta$
	\begin{equation}
	\langle m_z \rangle =\cos\Theta = \epsilon \delta V + \frac{ \alpha \epsilon^2 k  \sin^4\Theta \Omega_F }{ 2V(1 + \alpha^2 + \delta \alpha \epsilon k \sin^2\Theta \, \Omega_F)^2 } \label{Eq_cond_A0_Theta_main}
	\end{equation}
\end{subequations}
Equation~(\ref{Eq_cond_A0_Theta_main}) is valid for $-1 \le \epsilon V \le 1$; if $|\epsilon V| > 1$, then $m_z = {\rm sign}(V)$. %, where ${\rm sign}(x) = 1$ if $x\ge 0$ and $-1$ if $x<0$.
The right hand side of this equation consists of the quasiparticle current term (first term), which comes from the $0$-th order contribution to the slow motion and creates constant magnetic field in $m_z=+1$ direction. The second term appears because of the superconducting current oscillations, which comes from the $1$-th order contribution and generates a small adding to the reorientation of the magnetic moment. This term represents the Kapiza pendulum effect \cite{smrbb-epl_18}, since if the magnetic system is analogy of the pendulum in mechanics then the term $\epsilon\sin(V t - k m_{z})$ in the effective field is the oscillating force of the suspension point of this pendulum and the second term in the right hand side of (\ref{Eq_cond_A0_Theta_main}) is the contribution to the motion of the stable point from this force. Generally, the oscillating magnetic field from the second term creates two symmetric stable points (for example $m_z=\pm1$ for the complete reorientation). Because of the constant magnetic field generated by quasiparticle current in our setup there is only one stable point. 

\begin{figure}[t]
	\centering
	\includegraphics[width=4.5cm]{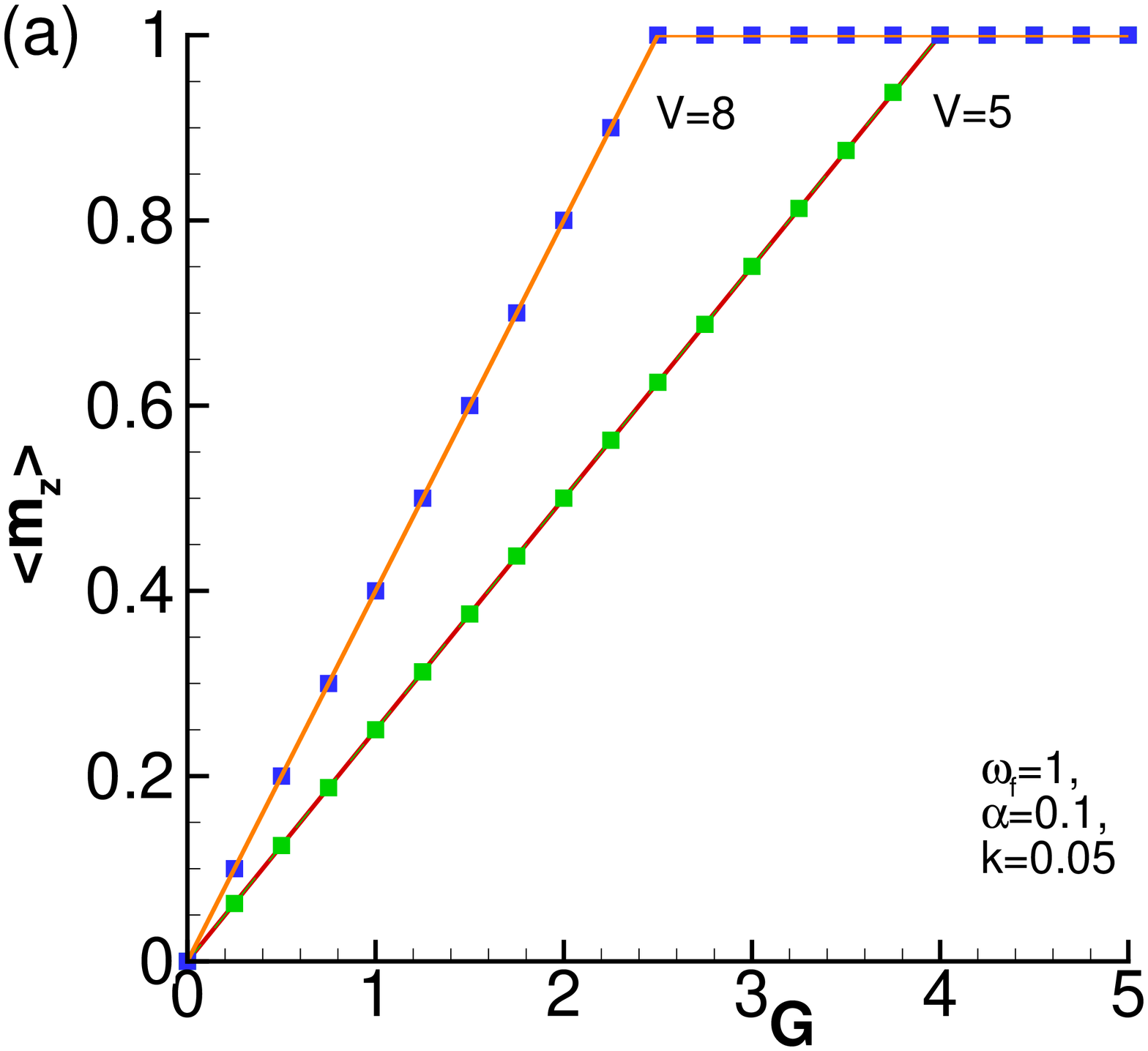}\includegraphics[width=4.5cm]{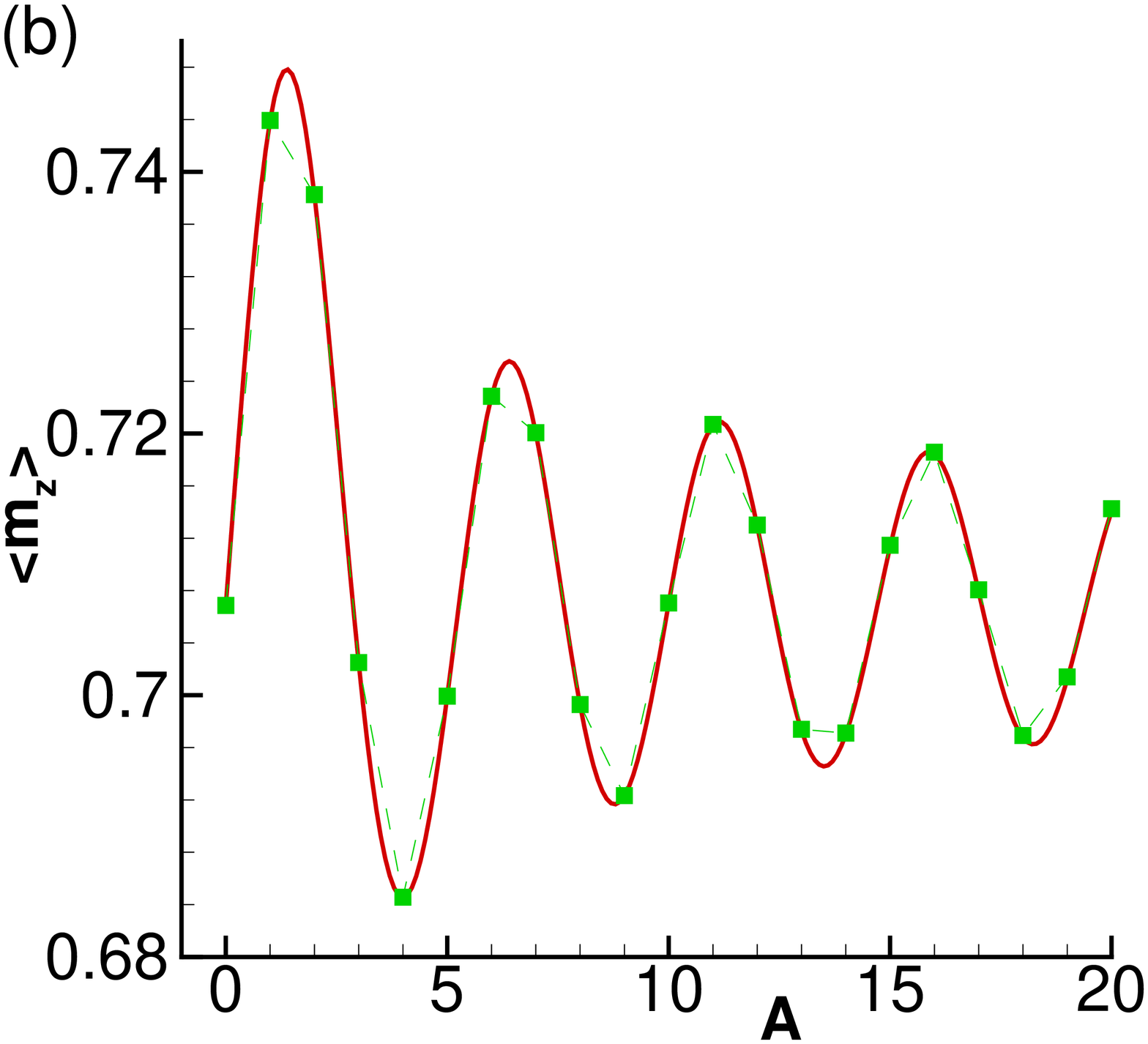}
	\includegraphics[width=4.5cm]{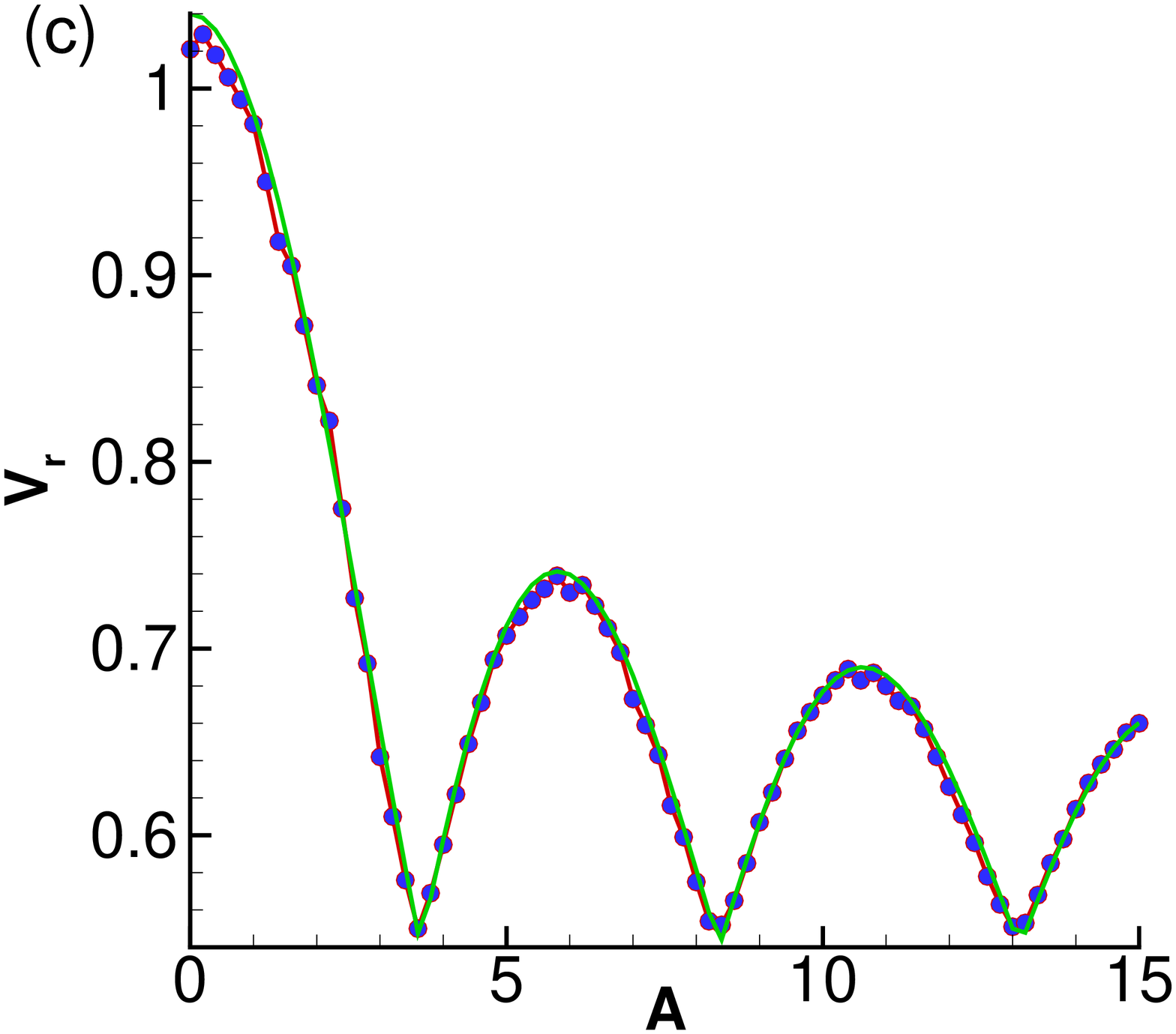}
	% I1(V,A,Omega).jpg: 2837x1679 px, 72dpi, 100.08x59.23 cm, bb=0 0 2837 1679
	\caption{(color online) (a) The average values of $m_z$ versus the Josephson-to-magnetic energy ratio $G$ for the two indicated voltages. The symbols show the values obtained by the numerical calculation of eq. (\ref{LLG_components}). The lines show analytical results obtained from eqs. (\ref{Eq_cond_A0_main}). (b) The average value of the magnetic moment component $m_z$ as a function of amplitude of external drive $A$ at $\alpha = 0.1$, $G = 3\pi$, $k = 0.1$, $\Omega_F = 0.001$, $V=0.75$, $\Omega = 0.75$, $m_0=-V/\Omega=-1$. The symbols indicate the average values obtained by numerical calculation of Eq. (\ref{LLG_components}). The lines indicate the analytical results  obtained from eqs. (\ref{Eq_cond_C2_Theta_main}) and (\ref{Eq_cond_A0_Phi_main}). (c) Variation of the reorientation voltage with the external drive amplitude. The green line represents the fitting with Bessel function.}
	\label{fig_I1_I2_I3}
\end{figure}

Results of numerical calculations of the averaged $m_z$ as a function of $G$ at two frequencies, presented in Fig.~\ref{fig_I1_I2_I3}(a), demonstrate the changes of stability position. The analytical dependence  calculated by (\ref{Eq_cond_A0_main}) shows an excellent agreement with numerical data.
Our results explain the unusual frequency dependence of the reorientation in the $\varphi_0$ Josephson junction \cite{smrbb-epl_18}, where only the superconducting current in the effective field of the LLG equation was taken into account.
This discrepancy with the usual Kapitsa pendulum is due to the omission of the quasiparticle current in the effective field.
It can be clearly seen in (\ref{Eq_cond_A0_Theta_main}) that at $\delta=0$ (without quasiparticle current in the effective field \cite{smrbb-epl_18}) only the second term contributes, which is proportional to $1/V$, but at $\delta=1$ the first term in the equation, which is proportional to $V$, increases much faster.
In our considerations, both currents are included in the effective field, therefore, the frequency dependence shown in Fig.\ref{fig:case1}(b) coincide with the usual Kapitsa pendulum features. Note also that the value of $G$ for the complete reorientation, which indicates the stabilization of the magnetic moment dynamics at $\delta=1$ is much smaller than at $\delta=0$ \cite{smrbb-epl_18}. 

%\paragraph*{Stability under external drive and the zeroth order resonances.}
\section{Stability under external drive and the zeroth order resonances}

If we have a zeroth order resonance and the conditions $V+m_0\Omega = 0$ (see Section~\ref{subsubsec_stability_m0}) are satisfied. Then from ~(\ref{dTdP0}) and $\dot\Theta = \dot\Phi = 0$ we obtain $\Phi = \pi/2$ or $3\pi/2$ and an equation for $\Theta$:
\begin{equation} \label{Eq_cond_C2_Theta_main}
	\cos\Theta = \epsilon \delta V - \epsilon \text{sign}^{m_0}(m_0) J_{m_0}\left(\frac{A}{\Omega}\right)\sin(k \cos\Theta ) .
\end{equation}
The right hand side of this equation again consists of two terms: quasiparticle current term (first term) and finite average superconducting current term. Both of these terms come from the $0$-th order contribution to the slow motion and create a constant magnetic field in $m_z=+1$ direction. The finite average superconducting current occurs from the condition $V+m_0\Omega = 0$ which is the Shapiro step condition in the JJ. In this case oscillations of the junction are locked by the external periodic drive and as a result a finite average superconducting current appears. 

Based on (\ref{Eq_cond_C2_Theta_main}), we calculate the average value of magnetic moment component $m_z$ as a function of amplitude of external drive $A$. The results are shown in Fig.~\ref{fig_I1_I2_I3}(b) and we see a very good agreement with the direct numerical calculations. Note that the average $m_z$ as a function of $A$ demonstrates the Bessel behavior.

The applied external periodic drive also affects the voltage of complete reorientation $V_{r}$, which indicates the stabilization of the magnetic moment dynamics (see Fig.~\ref{1}(b)). This effect is demonstrated in  Fig. \ref{fig_I1_I2_I3}(c), where the results of numerical  calculations of $V_{r}$ as a function of external drive amplitude $A$ are presented (symbols). $V_{r}$ as a function of $\Omega$ is similar (is not shown here).

The numerical data are well ﬁtted by the Bessel function. At chosen parameters $G=3\pi, ~\alpha=0.1, ~k=0.1, ~\Omega_F=0.5$ and $\Omega=0.8$ the data present a good agreement with Bessel dependence $0.5J(A/\Omega)+0.54$, shown by green solid line. Note, that this effect can also be seen from (\ref{Eq_cond_C2_Theta_main}) at $\cos\Theta =1 $.

It is important to discuss some limiting cases in the framework of our model. For example, what would happen in the purely normal state ($I_c=0$) and in the case of an isolating interlayer ($R=\infty$). In the first case, there would be two terms in the z-component of the effective field (\ref{Effective_Field_comp}), one periodic, corresponding to the external periodic drive, and another one constant. As a result, one would get something similar to (\ref{Eq_cond_A0_Theta_main}) for the reorientation: there would be the $0$-th order contribution which comes from the constant term and the $1$-th order contribution which comes from the periodic term. In the second case, there would the periodic term in the z-component of the effective field only, which is related to the external periodic drive. So, the $0$-th order term would not appear and the $1$-th and higher order terms would contribute to the reorientation of the magnetic moment.

%\paragraph*{Conclusions.}
\section{Conclusions}

We demonstrated the reorientation effects in a system of nanomagnet coupled to the Josephson junction under the influence of external periodic drive.
It was shown that the dependence of the average value of the magnetic moment component $m_z$ on the amplitude of external drive $A$ is described by the Bessel function. In comparison with the earlier works, where the magnetic moment flips and the reorientation by the dc voltage or current was reported \cite{kulik1966zh, bulaevskii1977pis, thirion2nature, cc-prb_10, smrbb-epl_18, snrk-jetpl_19}, this result considerably expands the possibilities of controlling the magnetization dynamics. It can be especially significant in the light of the fact that such a system could be used as qubit in quantum information processing. We emphasize that this influence is orders of magnitude more pronounced when the Josephson frequency $V$ is equal to an integer number of external drive frequencies $\Omega$ (i.e. $V+m_0\Omega = 0$, where $m_0$ is negative integer). Otherwise, the influence of $A$ is very small and the Kapitsa pendulum-like  effects come mainly from the Josephson oscillations. We also showed that the voltage of complete reorientation $V_{r}$ depends as the Bessel function on the external drive amplitude. We obtained analytical expressions to describe the movement of the stability position in the $yz$ plane and had a very good agreement with the direct numerical simulations. It was shown that the magnetic field of the quasiparticle current determines the frequency dependence of magnetic moment's stable position. It also decreases the value of Josephson to magnetic energy ratio $G$ necessary for the complete reorientation. Therefore, the quasiparticle current magnetic field plays an important role for the reorientation problem.

The experimental verification of our work would involve preparing a voltage-biased JJ-nanomagnet system with sufficiently small values of $\epsilon$. We expect that a 2D thin-film niobium superconducting junction in the $y-z$ plane coupled to the nanomagnet could be a potential candidate for experimental realization. Note that the value of the Josephson energy in such a junction is $\epsilon_J \sim 2 \times 10^{-18} \; J$, while the resistance is $\sim 3 \; m \Omega$ and $\omega_c = 50 \; GHz$. So, the Josephson frequency is in the range of $GHz$ and the voltage in $\mu V$. The nanomagnet is assumed to have a radius of $7 - 30 \; nm$ in thickness with magnetic anisotropy constant of $K \sim 20\; kJ/m^3$ and a saturation magnetization of $1950\; kA/m$. The Josephson junction induced by an electromagnetic radiation of frequency around $1 GHz$. We also note that such experiments should also be possible with 1D junctions using nanowires with spin-orbit coupling \cite{rokhinson2012fractional, mourik2012signatures}. 
The coupling of such JJs with Majorana bound states to nanomagnets in the presence of a magnetic field may lead to new experimental signatures of such states \cite{ghosh2017magnetization}. 
We consider that the obtained results open a wide field of research and applications related to the possibility of reorientation  of nanomagnet's easy axis. Such a realization might play a crucial role in quantum information processing and spintronics.

%\paragraph*{Acknowledgements}

\begin{acknowledgments}
The reported study was funded by the RFBR research project 20-37-70056, UEFISCDI project PN 19060101, Egypt-JINR research project for the year 2021, and Romania-JINR collaboration project 23/365/2021. Special thanks to JINR(Russia) and Bibliotheca Alexandrina (Egypt) HPC for the calculating servers.
\end{acknowledgments}

\appendix

\begin{widetext}

\section{Basic equations} \label{sec_BE}

We start from Eqs.~(\ref{LLG_components}), which we rewrite explicitly on components, together with the functions and parameters involved:
\begin{subequations} \label{derivs_mxmymz}
\begin{eqnarray}
		&& \frac{d m_x}{d t} = \frac{\Omega_F}{1+\alpha^2} [ \alpha h_x (m_y^2 + m_z^2) + h_y (m_z - \alpha m_x m_y) - h_z (\alpha m_x m_z + m_y) ] , \label{deriv_mx} \\
			&& \frac{d m_y}{d t} = \frac{\Omega_F}{1+\alpha^2} [ - h_x (m_z + \alpha m_x m_y) + \alpha h_y (m_x^2 + m_z^2) + h_z (m_x - \alpha m_y m_z)] , \label{deriv_my} \\
			&& \frac{d m_z}{d t} = \frac{\Omega_F}{(1+\alpha^2) D} [ h_x (m_y - \alpha m_x m_z) - h_y (m_x + \alpha m_y m_z) + \alpha \tilde h_z (m_x^2 + m_y^2)] , \label{deriv_mz} \\
			&& h_x = 0, \quad h_y = m_y, \quad h_z = \tilde h_z - \delta \epsilon k \dot m_z , \quad
			D = 1 + \frac{\Omega_F \delta \alpha \epsilon k}{1+\alpha^2} (m_x^2 + m_y^2)
			\equiv  1 + \frac{\Omega_F \delta \alpha \epsilon k}{1+\alpha^2} \sin^2\theta ,
			\label{def_D} \\
			&& \tilde h_z(t) = \epsilon \left\{ \sin\left[ V t - k m_z + \frac{A}{\Omega} \sin(\Omega t) \right] + \delta \left[V + A \cos(\Omega t) \right] \right\}
			\equiv \epsilon \delta V + \tilde g_z(t) ,
			\label{def_thz}
			%
			%			&& \alpha = 0.1, \ \Omega_F = 1, k = 0.01, \epsilon = 2\pi , \ A \in [1, 30], \ \beta_c = 0.25
			%			\label{val_params}
		\end{eqnarray}
	\end{subequations}
	where the projections of the magnetization and their time derivatives are
	\begin{subequations} \label{mxmymz_pol_coord_etal}
		\begin{eqnarray}
			&& m_x = \sin\theta \cos\phi,\
			m_y = \sin\theta \sin\phi,\
			m_z = \cos\theta , \label{mxmymz_pol_coord} \\
			&& \dot m_x = \dot \theta \cos(\theta) \cos(\phi) - \dot \phi \sin(\theta) \sin(\phi), \quad
			\dot m_y = \dot \theta \cos(\theta) \sin(\phi) + \dot \phi \sin(\theta) \cos(\phi), \quad
			\dot m_z = - \dot\theta \sin\theta . \label{tder}
		\end{eqnarray}
	\end{subequations}
	By plugging Eqs.~(\ref{mxmymz_pol_coord_etal}) into~(\ref{derivs_mxmymz}), we obtain
	\begin{subequations} \label{derivs_mxmymz_pol}
		\begin{eqnarray}
			\dot \theta \cos(\theta) \cos(\phi) - \dot \phi \sin(\theta) \sin(\phi) % \nonumber \\
			&=& \frac{\Omega_F}{1+\alpha^2} \Big[ \sin(\phi) (\cos\theta - \alpha \sin^{2}\theta \cos\phi \sin\phi) - (\tilde h_z - \delta \epsilon k \dot m_z) (\alpha \cos\phi \cos\theta +  \sin\phi) \Big] \nonumber \\
			&& \times \sin\theta
			\label{deriv_pc1_pol} \\
			%%%%%%
			\dot \theta \cos(\theta) \sin(\phi) + \dot \phi \sin(\theta) \cos(\phi) % \nonumber \\
			&=& \frac{\Omega_F}{1+\alpha^2} \Big[ \alpha \sin\phi (\sin^2\theta \cos^2\phi + \cos^2\theta) + (\tilde h_z - \delta \epsilon k \dot m_z) (\cos\phi - \alpha \sin\phi \cos\theta) \Big] \nonumber \\
			&& \times \sin\theta
			\label{deriv_pc2_pol} \\
			%%%%%%%%%%%%%%%%
			\dot m_z = - \dot\theta \sin(\theta)
			&=& \frac{\sin^2\theta \,\Omega_F}{ 1 + \alpha^2 + \delta \alpha \epsilon k \sin^2\theta \, \Omega_F} \left[ \alpha \tilde h_z - \sin\phi (\cos\phi + \alpha \cos\theta \sin\phi) \right]  \label{deriv_pc3_pol}
		\end{eqnarray}
		For $\theta \ne 0,\pi$, Eq.~(\ref{deriv_pc3_pol}) implies
		\begin{equation}
			\dot\theta = - \frac{\sin\theta \, \Omega_F}{ 1 + \alpha^2 + \delta \alpha \epsilon k \sin^2\theta \, \Omega_F}  \Big[ \alpha \tilde h_z - \sin\phi ( \cos\phi + \alpha \cos\theta \sin\phi) \Big] .
			\label{deriv_theta}
		\end{equation}
	\end{subequations}
	Equation~(\ref{deriv_theta}) may be extended by continuity from $(0,\pi)$ to the closed interval $[0,\pi]$, by $\dot{\theta}(0) = \dot{\theta}(\pi) = 0$.
	%Generally, Eqs.~(\ref{derivs_mxmymz_pol}) are valid only if $\sin\theta \ne 0$.
	%However, due to continuity, we observe that $\dot\theta \to 0$ as $\theta \to 0$, so Eq.~(\ref{deriv_theta}) is valid also if $\sin\theta = 0$.

	Equations (\ref{deriv_pc1_pol}) and (\ref{deriv_pc2_pol}) are equivalent and they lead to
	\begin{subequations} \label{eq_dot_phi_equiv}
		\begin{eqnarray}
			\dot \phi &=&
			- \frac{\Omega_F}{(1+\alpha^2)} \left\{ \frac{ - \sin\phi ( \cos\phi + \alpha \cos\theta \sin\phi) + \alpha \tilde h_z }{ \left( 1 + \frac{\Omega_F \delta \alpha \epsilon k}{1+\alpha^2} \sin^2\theta \right) } \cos(\theta) \cot \phi + \cos\theta - \alpha \sin^{2}\theta \cos\phi \sin\phi \right.
			\nonumber \\
			&& \left. - \left[ \tilde h_z - \delta \epsilon k \frac{\Omega_F}{(1+\alpha^2)} \frac{ - \sin\phi (\cos\phi + \alpha \cos\theta \sin\phi) + \alpha \tilde h_z }{ 1 + \frac{\Omega_F \delta \alpha \epsilon k}{1+\alpha^2} \sin^2\theta } \sin^2\theta \right] (1 + \alpha \cos\theta \cot \phi) \right\}
			\label{eq_dot_phi1}
		\end{eqnarray}
		and
		\begin{eqnarray}
			\dot \phi &=& \frac{\Omega_F}{1+\alpha^2} \left\{ \frac{ - \sin\phi ( \cos\phi + \alpha \cos\theta \sin\phi) + \alpha \tilde h_z }{ \left( 1 + \frac{\Omega_F \delta \alpha \epsilon k}{1+\alpha^2} \sin^2\theta \right) } \cos\theta \tan \phi + \alpha \tan\phi (\sin^2\theta \cos^2\phi + \cos^2\theta) \right. \nonumber \\
			&& \left. + \left[ \tilde h_z - \delta \epsilon k \frac{\Omega_F}{(1+\alpha^2)} \frac{ - \sin\phi (\cos\phi + \alpha \cos\theta \sin\phi) + \alpha \tilde h_z }{ 1 + \frac{\Omega_F \delta \alpha \epsilon k}{1+\alpha^2} \sin^2\theta } \sin^2\theta \right] (1 - \alpha \cos\theta \tan\phi) \right\} ,
			\label{eq_dot_phi2}
		\end{eqnarray}
	\end{subequations}
	respectively.
	Equations.~(\ref{eq_dot_phi_equiv}) above may be reduced to the same expression:
	\begin{eqnarray}
		\dot \phi &=& \frac{\Omega_F}{ 1 + \alpha^2 + \delta \alpha \epsilon k \sin^2\theta \, \Omega_F}  \left[ \tilde h_z
		- \Big( - \sin^2 \theta \cos\phi \delta \epsilon k \Omega_F + \sin\phi \cos\theta - \alpha \cos\phi \Big) \sin(\phi) \right] , \label{eq_dot_phi_simple}
	\end{eqnarray}
	so we have to solve a system of two coupled equations,  namely (\ref{deriv_theta}) and~(\ref{eq_dot_phi_simple}).

	\section{Effects similar to the Kapitsa pendulum} \label{sec_app_K}

If $V$ and $\Omega$ are much bigger than the oscillation frequency the nano-magnet without any external influence (Josephson junction and external radiation), a set of features specific to the Kapitsa pendulum emerge.
	To study this, we introduce the notations
	\begin{equation}
		\theta \equiv \Theta + \xi \qquad {\rm and} \qquad \phi \equiv \Phi + \zeta
		\label{defs_xi_zeta}
	\end{equation}
	in order to separate the "slow" and "fast" components pf the motion. Because $\xi$ and $\zeta$ are small, we may write
	\begin{subequations} \label{dot_xi_zeta}
		\begin{eqnarray}
			\dot\theta &=& \dot\Theta + \dot\xi \approx
			\frac{ \Omega_F }{ 1 + \alpha^2 + \alpha \delta \epsilon k \Omega_F \sin^2\Theta }
			\Bigg\{ \left( ( \alpha \sin\Phi \cos\Theta + \cos\Phi) \sin\Phi- \alpha \epsilon \delta V
			- \alpha \epsilon \sin\left[ V t - k \cos\Theta + \frac{A}{\Omega} \sin(\Omega t) \right] \right) \sin\Theta
			\nonumber \\
			&& + \left( \frac{- \cos\Phi \cos\Theta \Big(\sin^2\Theta  \alpha \delta \epsilon k \Omega_F  - \alpha^2 - 1 \Big) + 2 \alpha ( \alpha^2 + 1) \sin\Phi \cos^2\Theta - \alpha (\alpha \delta \epsilon k \Omega_F \sin^2\Theta + \alpha^2 + 1) \sin\Phi }{ \alpha \delta \epsilon k \Omega_F \sin^2\Theta + \alpha^2 + 1}
			\right. \nonumber \\
			&& \times \sin\Phi
			- \alpha \epsilon k \sin^2\Theta \cos\left[ V t - k \cos\Theta + \frac{A}{\Omega} \sin(\Omega t) \right]
			- \frac{ 1 + \alpha^2 - \alpha \delta \epsilon k \Omega_F \sin^2\Theta }{ 1 + \alpha^2 + \alpha \delta \epsilon k \Omega_F \sin^2\Theta }
			\Bigg\{ \alpha \epsilon \cos\Theta
			\nonumber \\
			&& \left. \times \sin \left[Vt - k\cos\Theta + \frac{A}{\Omega} \sin(\Omega t) \right] + \delta \epsilon \alpha V \cos \Theta + \delta \epsilon \alpha A \cos \Theta \cos(\Omega t)
			\Bigg\} \right) \xi
			\nonumber \\
			&& + (2 \cos\Phi  \sin\Phi  \cos\Theta \, \alpha + 2 \cos^2\Phi  - 1) \sin\Theta \, \zeta
			- \alpha \sin\Theta \epsilon \delta A  \cos(\Omega t)  \Bigg\} ,
			\label{dot_xi}\\
			%%%%%%%%%%%%%%%%%%%%%%%%
			\dot\phi &=& \dot\Phi + \dot\zeta \approx
			\frac{ \Omega_F }{ \Omega_F \alpha \delta \epsilon k \sin^2\Theta + \alpha^2 + 1} \Bigg( \Big[\epsilon \delta V + (\delta \epsilon k \Omega_F \sin^2\Theta \cos\Phi - \cos\Theta \sin\Phi + \alpha \cos\Phi) \sin\Phi \Big]
			\nonumber \\
			&& + \epsilon \sin\left[ V t - k m_z + \frac{A}{\Omega} \sin(\Omega t) \right]
			+ \left( \frac{-2 \delta\alpha\epsilon k\Omega_{F}\cos\Theta}{1+\alpha^{2}+\delta \alpha\epsilon k\Omega_{F} \sin^2\Theta}
			\left\{ \delta \epsilon \Big[ V+A\cos(\Omega t)  \Big]
			\right. \right. \nonumber \\
			&& \left. +
			\Big( \delta\alpha\epsilon k\Omega_{F} \sin^2\Theta \cos\Phi - \sin\Phi \cos\Theta +\alpha \cos\Phi \Big) \sin\Phi  \right\}
			+ \Big( 2 \delta\alpha\epsilon k\Omega_{F} \cos\Phi \cos\Theta + \sin\Phi \Big) \sin\Phi \nonumber \\
			&& \left. + \epsilon k\cos \left( Vt- k\cos\Theta + {\frac {A\sin(\Omega t) }{\Omega}} \right) - \frac{2 \delta\alpha\epsilon^{2}k\Omega_{F} \cos\Theta}{1 + \alpha^{2}+\delta\alpha\epsilon k\Omega_{F} \sin^2\Theta} \sin\left( Vt-k\cos\Theta + \frac{A\sin(\Omega t)}{\Omega} \right)\right) \sin\Theta \, \xi \nonumber \\
			&& + \Big[ (2\cos^2\Phi  - 1) (k \delta \epsilon \Omega_F \sin^2\Theta + \alpha) - 2 \cos\Phi  \cos\Theta  \sin\Phi \Big] \zeta
			+ \epsilon \delta A  \cos(\Omega t) \Bigg).
			\label{dot_zeta}
		\end{eqnarray}
	\end{subequations}
	The ``slow'' velocity $(\dot{\Theta},  \dot{\Phi})$ may be calculated from Eqs.~(\ref{dot_xi_zeta}) by averaging over the ``fast'' and small oscillations described by the velocity $(\dot{\xi}, \dot{\zeta})$.
	%	If the (angular) frequencies $V$ and $\Omega$ are much higher than the characteristic frequency of the system $\Omega_F$ (without coupling to the external radiation), then we can calculate the ``average velocities,'' denoted as $\dot\Theta$ and $\dot\Phi$.
	%
	In order  to do this, we further process Eqs.~(\ref{dot_xi_zeta}) by writing
	\begin{subequations} \label{dec_sin_cos}
		\begin{eqnarray}
			\sin\left[ V t - k m_z + \frac{A}{\Omega} \sin\left( \Omega t \right) \right]
			&=& \sin ( V t - k m_z ) \cos\left[ \frac{A}{\Omega} \sin\left( \Omega t \right) \right]
			+ \cos ( V t - k m_z ) \sin\left[ \frac{A}{\Omega} \sin\left( \Omega t \right) \right]
			%	\nonumber \\
			%	%
			%	&\equiv& g_1(t) + g_2(t)
			, \label{dec_sin1} \\
			%%%%%%%%%%%%%%%%
			\cos\left[ V t - k m_z + \frac{A}{\Omega} \sin\left( \Omega t \right) \right]
			&=& \cos ( V t - k m_z ) \cos\left[ \frac{A}{\Omega} \sin\left( \Omega t \right) \right]
			- \sin ( V t - k m_z ) \sin\left[ \frac{A}{\Omega} \sin\left( \Omega t \right) \right]
			%	\nonumber \\
			%	%
			%	&\equiv& h_1(t) + h_2(t)
			\label{dec_cos1}
		\end{eqnarray}
	\end{subequations}
	and then by using the expansions
	\begin{subequations} \label{exp_sin_cos_Bessel}
		\begin{eqnarray}
			\sin[z \sin(\gamma)] &=& 2 \sum_{k=0}^{\infty} J_{2k+1}(z) \sin[(2k+1) \gamma] , \label{exp_sin_Bessel} \\
			\cos[z \sin(\gamma)] &=& J_0(z) + 2 \sum_{k=1}^{\infty} J_{2k}(z) \cos[(2k) \gamma] . \label{exp_cos_Bessel}
		\end{eqnarray}
	\end{subequations}
	Plugging Eqs.~(\ref{exp_sin_cos_Bessel}) into~(\ref{dec_sin_cos}), we obtain
	\begin{subequations} \label{exp_sin_cos_Bessel2}
		\begin{eqnarray}
			&& % g_1(t) + g_2(t)
			\sin\left[ V t - k m_z + \frac{A}{\Omega} \sin\left( \Omega t \right) \right] = \sum_{m=-\infty}^{\infty} \text{sign}^m(m) J_{|m|}\left(\frac{A}{\Omega}\right)
			\sin \Big[ (V + m \Omega) t - k m_z \Big] \label{def_g1pg2} \\
			&& % h_1(t) + h_2(t)
			\cos\left[ V t - k m_z + \frac{A}{\Omega} \sin\left( \Omega t \right) \right] = \sum_{m=-\infty}^{\infty} \text{sign}^m(m) J_{|m|}\left(\frac{A}{\Omega}\right)
			\cos \Big[ (V + m \Omega) t - k m_z \Big] ,
			\label{def_h1ph2}
		\end{eqnarray}
	\end{subequations}
	which may then be used in Eqs.~(\ref{dot_xi_zeta}).
	It is straightforward to verify from the right hand sides of Eqs.~(\ref{exp_sin_cos_Bessel2}) that $\sin\left[ V t - k m_z + \frac{A}{\Omega} \sin\left( \Omega t \right) \right]_{A=0} = \sin\left[ V t - k m_z \right]$ and $\cos\left[ V t - k m_z + \frac{A}{\Omega} \sin\left( \Omega t \right) \right]_{A=0} = \cos\left[ V t - k m_z \right]$.

	\subsection{Solutions by iterations} \label{subsec_iter}

	Equations~(\ref{dot_xi_zeta}) may be solved by iterations.
	For this, we define in the lowest (zeroth) order the velocity of the slow motion
	\begin{subequations} \label{dot_Theta1_Phi1}
		\begin{eqnarray}
			\dot\Theta_0 &\equiv& \frac{ \sin\Theta \, \Omega_F }{  1 + \alpha^2 + \delta \alpha \epsilon k \sin^2\Theta \, \Omega_F }
			\left[ ( \alpha \sin\Phi \cos\Theta + \cos\Phi) \sin\Phi- \alpha \epsilon \delta V
			\right] ,
			\label{dot_Theta1}\\
			%%%%%%%%%%%%%%%%%%%%%%%%
			\dot\Phi_0 &\equiv& \frac{ \Omega_F }{  1 + \alpha^2 + \delta \alpha \epsilon k \sin^2\Theta \, \Omega_F} \left[ \epsilon \delta V + (\delta \epsilon k \Omega_F \sin^2\Theta \cos\Phi - \cos\Theta \sin\Phi + \alpha \cos\Phi) \sin\Phi
			\right] ,
			\label{dot_Phi1}
		\end{eqnarray}
	\end{subequations}
	and of the fast motion,
	\begin{subequations} \label{dot_xi_zeta_it0}
		\begin{eqnarray}
			\dot\xi_0 &=& - \alpha \sin\Theta \dot\zeta_0 ,
			\label{dot_xi_it0}\\
			%%%%%%%%%%%%%%%%%%%%%%%%
			\dot\zeta_0 &\equiv&
			\frac{ \epsilon \Omega_F }{  1 + \alpha^2 + \delta \alpha \epsilon k \sin^2\Theta \, \Omega_F}
			\left\{ \sin\left[ V t - k m_z + \frac{A}{\Omega} \sin\left( \Omega t \right) \right]
			+ \delta A \cos(\Omega t) \right\}  \label{dot_zeta_it0} \\
			&=& \frac{ \epsilon \Omega_F }{ 1 + \alpha^2 + \delta \alpha \epsilon k \sin^2\Theta \, \Omega_F}
			\left\{ \sum_{m=-\infty}^{\infty} \text{sign}^m(m) J_{|m|}\left(\frac{A}{\Omega}\right)
			\sin \left[ (V + m \Omega) t - k m_z \right]
			+ \delta A  \cos(\Omega t) \right\} . \nonumber
		\end{eqnarray}
	\end{subequations}
	Integrating over time we obtain $\xi_0 = - \alpha \sin\Theta \, \zeta_0$ and
	%
	%\begin{subequations} \label{xi_zeta_it0}
		\begin{equation}
			\zeta_0 =
			\frac{ \epsilon \Omega_F }{ \Omega_F \alpha \delta \epsilon k \sin^2\Theta + \alpha^2 + 1}
			\left\{ \sum_{m=-\infty}^{\infty} - \text{sign}^m(m) J_{|m|}\left(\frac{A}{\Omega}\right)
			\frac{\cos \Big[ (V + m \Omega) t - k m_z \Big]}{V + m \Omega}
			+ \delta A  \frac{\sin(\Omega t)}{\Omega}  \right\} . \label{zeta_it0}
		\end{equation}
		%\end{subequations}
	%%
	Plugging Eqs.~(\ref{dot_xi_zeta_it0})  and (\ref{zeta_it0}) into~(\ref{dot_xi_zeta}), we obtain the next order iteration
	\begin{subequations} \label{dot_xi_zeta_it1}
		\begin{eqnarray}
			\dot\xi_1
			%%%%%%%%%%%%%%%%%%
			&=& \frac{ \Omega_F }{ 1 + \alpha^2 + \alpha \delta \epsilon k \Omega_F \sin^2\Theta }
			\Bigg( - \alpha \epsilon \sin\Theta \sum_{m=-\infty}^{\infty} \text{sign}^m(m) J_{|m|}\left(\frac{A}{\Omega}\right)
			\sin \Big[ (V + m \Omega) t - k m_z \Big]
			\nonumber \\
			&& - \left[ \frac{- \cos\Phi \cos\Theta \Big(\sin^2\Theta  \alpha \delta \epsilon k \Omega_F  - \alpha^2 - 1 \Big) + 2 \alpha ( \alpha^2 + 1) \sin\Phi \cos^2\Theta - \alpha (\alpha \delta \epsilon k \Omega_F \sin^2\Theta + \alpha^2 + 1) \sin\Phi }{ \alpha \delta \epsilon k \Omega_F \sin^2\Theta + \alpha^2 + 1}
			\right. \nonumber \\
			&& \left. + \frac{ - 1 - \alpha^2 + \alpha \delta \epsilon k \Omega_F \sin^2\Theta }{ 1 + \alpha^2 + \alpha \delta \epsilon k \Omega_F \sin^2\Theta } \delta \epsilon \alpha V \cos \Theta \right]
			\frac{ \alpha \sin\Phi \sin\Theta \epsilon \Omega_F }{ \Omega_F \alpha \delta \epsilon k \sin^2\Theta + \alpha^2 + 1}
			\Bigg\{ \sum_{m=-\infty}^{\infty} - \text{sign}^m(m) J_{|m|}\left(\frac{A}{\Omega}\right)
			\nonumber \\
			&& \times \frac{\cos \Big[ (V + m \Omega) t - k m_z \Big]}{V + m \Omega} + \delta A  \frac{\sin(\Omega t)}{\Omega}  \Bigg\} \nonumber \\
			&& + \frac{ \alpha^2 \epsilon^2 k \sin^3\Theta \Omega_F }{ \Omega_F \alpha \delta \epsilon k \sin^2\Theta + \alpha^2 + 1}
			\left( \sum_{m=-\infty}^{\infty} \text{sign}^m(m) J_{|m|}\left(\frac{A}{\Omega}\right)
			\cos \bigg\{ [ V + m \Omega ] t - k m_z \bigg\} \right) \nonumber \\
			&& \times \Bigg\{ \sum_{m=-\infty}^{\infty} - \text{sign}^m(m) J_{|m|}\left(\frac{A}{\Omega}\right)
			\frac{\cos \Big[ (V + m \Omega) t - k m_z \Big]}{V + m \Omega}
			+ \delta A \frac{\sin(\Omega t)}{\Omega}  \Bigg\} \nonumber \\
			&& + \frac{ -\alpha \epsilon \sin\Theta \Omega_F }{ \Omega_F \alpha \delta \epsilon k \sin^2\Theta + \alpha^2 + 1}
			\Bigg\{ \sum_{m=-\infty}^{\infty} - \text{sign}^m(m) J_{|m|}\left(\frac{A}{\Omega}\right)
			\frac{\cos \Big[ (V + m \Omega) t - k m_z \Big]}{V + m \Omega} \nonumber \\
			&&
			+ \delta A \bigg[ \frac{\sin(\Omega t)}{\Omega} + \beta_c \cos(\Omega t) \bigg] \Bigg\}
			\frac{ - 1 - \alpha^2 + \alpha \delta \epsilon k \Omega_F \sin^2\Theta }{ 1 + \alpha^2 + \alpha \delta \epsilon k \Omega_F \sin^2\Theta }
			\Bigg\{ \alpha \epsilon \cos\Theta
			\nonumber \\
			&& \times \sum_{m=-\infty}^{\infty} \text{sign}^m(m) J_{|m|}\left(\frac{A}{\Omega}\right)
			\sin \Big[ (V + m \Omega) t - k m_z \Big] + \delta \epsilon \alpha A \cos \Theta \cos(\Omega t)
			\Bigg\} \nonumber \\
			&& + (2 \cos\Phi  \sin\Phi  \cos\Theta  \alpha + 2 \cos^2\Phi  - 1)
			\frac{ \sin\Theta \epsilon \Omega_F }{ \Omega_F \alpha \delta \epsilon k \sin^2\Theta + \alpha^2 + 1} \nonumber \\
			&& \times \Bigg\{ \sum_{m=-\infty}^{\infty} - \text{sign}^m(m) J_{|m|}\left(\frac{A}{\Omega}\right)
			\frac{\cos \Big[ (V + m \Omega) t - k m_z \Big]}{V + m \Omega}
			+ \delta A  \frac{\sin(\Omega t)}{\Omega}  \Bigg\}
			- \alpha \sin\Theta \epsilon \delta A  \cos(\Omega t)  \Bigg)
			\label{dot_xi_it1}
		\end{eqnarray}
		and
		\begin{eqnarray}
			&& \dot\zeta_1 =
			\frac{ \Omega_F }{ \Omega_F \alpha \delta \epsilon k \sin^2\Theta + \alpha^2 + 1} \Bigg(
			\epsilon \sum_{m=-\infty}^{\infty} \text{sign}^m(m) J_{|m|}\left(\frac{A}{\Omega}\right)
			\sin \Big[ (V + m \Omega) t - k m_z \Big]
			\nonumber \\
			&& - \frac{ \epsilon \alpha \sin^2\Theta \Omega_F }{ \Omega_F \alpha \delta \epsilon k \sin^2\Theta + \alpha^2 + 1}
			\left\{ -2 \frac{\delta\alpha\epsilon k\Omega_{F}\cos\Theta }{1+{\alpha}^{2}+\delta \alpha\epsilon k\Omega_{F} \sin^2\Theta}
			\Big[ \delta \epsilon  V+A\cos(\Omega t)
			\right. \nonumber \\
			&& +
			\Big( \delta\alpha\epsilon k\Omega_{F} \sin^2\Theta \cos\Phi - \sin\Phi \cos\Theta +\alpha \cos\Phi \Big) \sin\Phi  \Big]
			+ \Big( 2 \delta\alpha\epsilon k\Omega_{F} \cos\Phi \cos\Theta + \sin\Phi \Big) \sin\Phi \nonumber \\
			&& + \epsilon k \sum_{m=-\infty}^{\infty} \text{sign}^m(m) J_{|m|}\left(\frac{A}{\Omega}\right)
			\cos \Big[ ( V + m \Omega ) t - k m_z \Big]
			-2 \frac {\delta\alpha\epsilon^{2}k\Omega_{F}
				\cos(\Theta) }{1+{\alpha}^{2}+\delta\alpha\epsilon k\Omega_{F} \sin^2\Theta}
			\nonumber \\
			&& \left. \times \sum_{m=-\infty}^{\infty} \text{sign}^m(m) J_{|m|}\left(\frac{A}{\Omega}\right)
			\sin \Big[ (V + m \Omega) t - k m_z \Big]
			\right\}
			\nonumber \\
			&& \times \Bigg\{ \sum_{m=-\infty}^{\infty} - \text{sign}^m(m) J_{|m|}\left(\frac{A}{\Omega}\right)
			\frac{\cos \Big[ (V + m \Omega) t - k m_z \Big]}{V + m \Omega}
			+ \delta A \frac{\sin(\Omega t)}{\Omega}  \Bigg\} \nonumber \\
			&& + \Big[ (2\cos^2\Phi  - 1) (k \delta \epsilon \Omega_F \sin^2\Theta + \alpha) - 2 \cos\Phi  \cos\Theta  \sin\Phi \Big]
			\frac{ \epsilon \Omega_F }{ \Omega_F \alpha \delta \epsilon k \sin^2\Theta + \alpha^2 + 1} \nonumber \\
			&& \times \Bigg\{ \sum_{m=-\infty}^{\infty} - \text{sign}^m(m) J_{|m|}\left(\frac{A}{\Omega}\right)
			\frac{\cos \Big[ (V + m \Omega) t - k m_z \Big]}{V + m \Omega}
			+ \delta A  \frac{\sin(\Omega t)}{\Omega}  \Bigg\}
			+ \epsilon \delta A  \cos(\Omega t) \Bigg)
			\label{dot_zeta_it1}
		\end{eqnarray}
	\end{subequations}
	In equations ~(\ref{dot_xi_zeta_it1}) we have terms of the type
	\begin{equation} \label{t_dep_terms}
		\sin \Big[ (V + m \Omega) t - k m_z \Big] ,
		\quad
		\cos \Big[ (V + m \Omega) t - k m_z \Big] ,
		\quad
		\sin(\Omega t) ,
		\quad
		\cos(\Omega t) ,
	\end{equation}
	and products between them.
	The products between the trigonometric functions lead to functions of different arguments:
	\begin{subequations} \label{prod_trigs}
		\begin{eqnarray}
			&& \sin^2 \Big[ (V + m \Omega) t - k m_z \Big] = \frac{1 - \cos[2(V + m \Omega) t - 2k m_z]}{2} , \label{prod_sin2} \\
			&& \cos^2 \Big[ (V + m \Omega) t - k m_z \Big] = \frac{1 + \cos[2(V + m \Omega) t - 2k m_z]}{2} , \label{prod_cos2} \\
			&& \sin\Big[ (V + m \Omega) t - k m_z \Big] \cos\Big[ (V + m \Omega) t - k m_z \Big] = \frac{\sin[2(V + m \Omega) t - 2k m_z]}{2} , \label{prod_sin_cos} \\
			&& \sin^2(\Omega t) = \frac{1 - \cos(2\Omega t)}{2} , \quad
			\cos^2(\Omega t) = \frac{1 + \cos(2\Omega t)}{2} , \quad
			\sin(\Omega t) \cos(\Omega t) = \frac{\sin(2\Omega t)}{2} , \label{prod_sin2_Ot} \\
			&& \sin(\Omega t) \sin\Big[ (V + m \Omega) t - k m_z \Big] = \frac{\cos\{[V + (m-1) \Omega] t - k m_z\} - \cos\{[V + (m+1) \Omega] t - k m_z\}}{2} , \label{prod_sinO_sin} \\
			&& \sin(\Omega t) \cos\Big[ (V + m \Omega) t - k m_z \Big] = \frac{\sin\{[V + (m+1) \Omega] t - k m_z\} - \sin\{[V + (m-1) \Omega] t - k m_z\}}{2} , \label{prod_sinO_cos} \\
			&& \cos(\Omega t) \sin\Big[ (V + m \Omega) t - k m_z \Big] = \frac{\sin\{[V + (m+1) \Omega] t - k m_z\} + \sin\{[V + (m-1) \Omega] t - k m_z\}}{2} , \label{prod_cosO_sin} \\
			&& \cos(\Omega t) \cos\Big[ (V + m \Omega) t - k m_z \Big] = \frac{\cos\{[V + (m+1) \Omega] t - k m_z\} + \cos\{[V + (m-1) \Omega] t - k m_z\}}{2} . \label{prod_cosO_cos}
		\end{eqnarray}
	\end{subequations}
	Of all the $\sin$ and $\cos$ functions from~(\ref{prod_trigs}), only the ones that do not depend on time contribute to the averages $\langle \dot{\xi} \rangle$ and $\langle \dot{\zeta} \rangle$, and therefore to the slow motion velocity in the next order of approximation $(\dot{\Theta}_1, \dot{\Phi}_1)$.

	One can continue the expansion to higher and higher orders and, by the same procedure, one gets trigonometric functions of arguments containing $(nV+m\Omega)t$, where $n$ and $m$ are integers.
	In the order $n$ of calculation we get terms of the form $(n'V+m\Omega)t$, where $n' \le n+1$.
%	We also notice that such terms are proportional to $M_0(\Theta) M^p(\Theta)$ or even a smaller factor, where
%	%
%	\begin{equation}
%		M_0(\Theta) \equiv \frac{ \Omega_F }{ 1 + \alpha^2+\alpha \delta \epsilon k \sin^2\Theta\, \Omega_F } \approx \Omega_F
%		\quad {\rm and} \quad
%		%M(\Theta) \equiv \frac{ \alpha \epsilon^2 k \sin^2\Theta \, \Omega_F }{ 1 + \alpha^2+\alpha \delta \epsilon k \sin^2\Theta\, \Omega_F}
%		M(\Theta) \equiv \frac{ \alpha \epsilon^2 k \Omega_F }{ 1 + \alpha^2+\alpha \delta \epsilon k \sin^2\Theta\, \Omega_F} \approx \alpha \epsilon^2 k \Omega_F ,
%		\label{def_M0M}
%	\end{equation}
%	%

We also notice that such terms are proportional to $M_0^{n+1}(\Theta)$ (\ref{F_theta}) and since we assume that $\Omega_F$ is much smaller than the frequencies $V$ and $\Omega$, the trigonometric functions of arguments containing $(n'V+m\Omega)t$ get smaller as $n'$ increases.
For this reason, in our expansions we shall stop to the leading order or the next to the leading order of approximation, since the higher orders are too small.

	\subsubsection{Stability without external periodic drive} \label{subsubsec_A0}

	All the time dependent terms average to zero, so the non-zero contributions to averages come only from the terms which are independent of time.
	When $A = 0$, Eqs.~(\ref{dot_xi_zeta_it1}) get a much simpler form, which lead to the averages
	\begin{subequations} \label{dot_xi_zeta_it1_A0}
		\begin{eqnarray}
			%	\dot\xi_1
			%	%%%%%%%%%%%%%%%%%%
			%	&=& \frac{ \Omega_F }{ (\alpha^2 + 1) \left( 1 + \frac{\alpha \delta \epsilon k \Omega_F \sin^2\Theta}{\alpha^2 + 1} \right) }
			%	\Bigg( \frac{ - \alpha^2 \epsilon^2 k \sin^3\Theta \Omega_F }{ \Omega_F \alpha \delta \epsilon k \sin^2\Theta + \alpha^2 + 1}\frac{\cos^2(Vt - km_z)}{V} \Bigg)
			%
			\langle \dot\xi_1 \rangle &=& \frac{ - \alpha^2 \epsilon^2 k \sin^3\Theta \Omega_F^2 }{(1 + \alpha^2 + \Omega_F \alpha \delta \epsilon k \sin^2\Theta)^2}\frac{1}{2V}
			= \frac{ - \alpha M(\Theta) \Omega_F \sin^3\Theta }{2V (1 + \alpha^2 + \Omega_F \alpha \delta \epsilon k \sin^2\Theta)} ,
			\label{dot_xi_it1_A0} \\
			%\end{eqnarray}
		%%
		%and
		%%
		%\begin{eqnarray}
			%	&& \dot\zeta_1 =
			%	\frac{ \Omega_F }{ \Omega_F \alpha \delta \epsilon k \sin^2\Theta + \alpha^2 + 1} \Bigg(
			%	\epsilon \sin(Vt - k m_z)
			%	+ \frac{ \epsilon \alpha \sin^2\Theta \Omega_F }{ \Omega_F \alpha \delta \epsilon k \sin^2\Theta + \alpha^2 + 1}
			%	\left\{ \frac{-2 \delta\alpha\epsilon k\Omega_{F}\cos\Theta }{1+{\alpha}^{2}+\delta \alpha\epsilon k\Omega_{F} \sin^2\Theta}
			%	\bigg[ \delta \epsilon V
			%	\right. \nonumber \\
			%	%
			%	&& +
			%	\Big( \delta\alpha\epsilon k\Omega_{F} \sin^2\Theta \cos\Phi - \sin\Phi \cos\Theta +\alpha \cos\Phi \Big) \sin\Phi  \bigg]
			%	+ \Big( 2 \delta\alpha\epsilon k\Omega_{F} \cos\Phi \cos\Theta + \sin\Phi \Big) \sin\Phi \nonumber \\
			%	%
			%	&&  \left. + \epsilon k \cos( V t - k m_z)
			%	-\frac{2 \delta\alpha\epsilon^{2}k\Omega_{F} \cos(\Theta)}{1 + \alpha^{2} + \delta\alpha\epsilon k\Omega_{F} \sin^2\Theta}
			%	\sin (Vt - k m_z)
			%	\right\} \frac{\cos(Vt - km_z)}{V} \nonumber \\
			%	%
			%	&& - \frac{ \epsilon \Omega_F }{ 1 + \alpha^2 + \Omega_F \alpha \delta \epsilon k \sin^2\Theta}
			%	\Big[ (2\cos^2\Phi  - 1) (k \delta \epsilon \Omega_F \sin^2\Theta + \alpha) - 2 \cos\Phi  \cos\Theta  \sin\Phi \Big]
			%	\frac{\cos(Vt - km_z)}{V} \Bigg)
			%
			\langle \dot\zeta_1 \rangle &=&
			\frac{ \alpha \epsilon^2 k \sin^2\Theta \Omega_F^2 }{ (1 + \alpha^2 + \Omega_F \alpha \delta \epsilon k \sin^2\Theta)^2} \frac{1}{2V}
			= \frac{ M(\Theta) \Omega_F \sin^2\Theta }{2V (1 + \alpha^2 + \Omega_F \alpha \delta \epsilon k \sin^2\Theta)} .
			\label{dot_zeta_it1_A0}
		\end{eqnarray}
	\end{subequations}
	Adding~(\ref{dot_xi_zeta_it1_A0}) to the zeroth order values~(\ref{dot_Theta1_Phi1}) we get
	\begin{subequations} \label{dot_Theta1_Phi1_c0}
		\begin{eqnarray}
			\dot\Theta_1 &\equiv& \frac{ \sin\Theta \, \Omega_F }{  1 + \alpha^2 + \delta \alpha \epsilon k \sin^2\Theta \, \Omega_F }
			\Bigg( ( \alpha \sin\Phi \cos\Theta + \cos\Phi) \sin\Phi- \alpha \epsilon \delta V
			- \frac{ \alpha M(\Theta) \Omega_F \sin^2\Theta }{2V (1 + \alpha^2 + \Omega_F \alpha \delta \epsilon k \sin^2\Theta)}
			\Bigg) ,
			\label{dot_Theta1_c0}\\
			%%%%%%%%%%%%%%%%%%%%%%%%
			\dot\Phi_1 &\equiv& \frac{ \Omega_F }{  1 + \alpha^2 + \delta \alpha \epsilon k \sin^2\Theta \, \Omega_F} \Bigg( (\delta \epsilon k \Omega_F \sin^2\Theta \cos\Phi - \cos\Theta \sin\Phi + \alpha \cos\Phi) \sin\Phi
			+  \epsilon \delta V
			\nonumber \\
			&& +\frac{ M(\Theta) \Omega_F \sin^2\Theta }{2V (1 + \alpha^2 + \Omega_F \alpha \delta \epsilon k \sin^2\Theta)}
			\Bigg) .
			\label{dot_Phi1_c0}
		\end{eqnarray}
	\end{subequations}
	From the equilibrium conditions $\dot\Theta_1 = \dot\Phi_1 = 0$ we obtain
	\begin{eqnarray}
		0 &=& ( 1+\alpha^2 + \alpha \delta \epsilon k \Omega_F \sin^2\Theta ) \cos\Phi ,
		\label{A0_eq_cond}
	\end{eqnarray}
	which implies,
	\begin{subequations} \label{Eq_cond_A0}
		\begin{equation}
			\Phi = \pi/2 \qquad {\rm or} \qquad \Phi = 3\pi/2 . \label{Eq_cond_A0_Phi}
		\end{equation}
		From equations~(\ref{dot_Theta1_Phi1}), and~(\ref{Eq_cond_A0_Phi}), we obtain
		\begin{equation}
			\cos\Theta = \epsilon \delta V + \frac{ M(\Theta) \Omega_F \sin^2\Theta }{2V (1 + \alpha^2 + \Omega_F \alpha \delta \epsilon k \sin^2\Theta)} , \label{Eq_cond_A0_Theta}
		\end{equation}
	\end{subequations}
	for $-1/\epsilon \le V \le 1/\epsilon$, otherwise, $\cos\Theta = \pm 1$.

	\subsubsection{Stability under external periodic drive and the zeroth order resonances} \label{subsubsec_stability_m0}

	Let us assume that there exists an integer $m_0 < 0$, such that $V+m_0\Omega = 0$ (that is, $m_0 = - V/\Omega \in \Z$).
	This will lead to a non-zero average contribution in the terms $\sin\left[ V t - k m_z + \frac{A}{\Omega} \sin\left( \Omega t \right) \right]$ and $\cos\left[ V t - k m_z + \frac{A}{\Omega} \sin\left( \Omega t \right) \right]$, which can be highlighted by writing the summations in the following manner:
	\begin{subequations} \label{g1pg2_h1ph2_c2}
		\begin{eqnarray}
			\sin\left[ V t - k m_z + \frac{A}{\Omega} \sin\left( \Omega t \right) \right] &=& - \text{sign}^{m_0}(m_0) J_{|m_0|}\left(\frac{A}{\Omega}\right) \sin ( k m_z )
			\nonumber \\
			&& + {\sum_{m=-\infty}^{\infty}}^{m\ne m_0} \text{sign}^m(m) J_{|m|}\left(\frac{A}{\Omega}\right) \sin \Big[ (V + m \Omega) t - k m_z \Big] , \label{def_g1pg2_c2} \\
			\cos\left[ V t - k m_z + \frac{A}{\Omega} \sin\left( \Omega t \right) \right] &=& \text{sign}^{m_0}(m_0) J_{|m_0|}\left(\frac{A}{\Omega}\right) \cos ( k m_z )
			\nonumber \\
			&& + {\sum_{m=-\infty}^{\infty}}^{m\ne m_0} \text{sign}^m(m) J_{|m|}\left(\frac{A}{\Omega}\right) \cos \bigg\{ [ V + m \Omega ] t - k m_z \bigg\} .
			\label{def_h1ph2_c2}
		\end{eqnarray}
	\end{subequations}
	This leads to a redefinition of the zeroth order quantities~(\ref{dot_Theta1_Phi1})-(\ref{zeta_it0}):
	\begin{subequations} \label{dot_Theta_Phi_0}
		\begin{eqnarray}
			\dot\Theta_0 &=&
			\frac{ \Omega_F }{  1+ \alpha^2+\alpha \delta \epsilon k \Omega_F \sin^2\Theta}
			\Bigg\{ ( \alpha \sin\Phi \cos\Theta + \cos\Phi) \sin\Theta \sin\Phi
			\nonumber \\
			&& - \alpha \epsilon \sin\Theta \Bigg[ \delta V - \text{sign}^{m_0}(m_0) J_{m_0}\left(\frac{A}{\Omega}\right) \sin ( k m_z )
			\Bigg] \Bigg\} ,
			\label{dot_Theta_0}\\
			%%%%%%%%%%%%%%%%%%%%%%%%
			\dot\Phi_0 &=&
			\frac{ \Omega_F }{  1+ \alpha^2+\alpha \delta \epsilon k \Omega_F \sin^2\Theta} \Bigg\{ (\delta \epsilon k \Omega_F \sin^2\Theta \cos\Phi - \cos\Theta \sin\Phi + \alpha \cos\Phi) \sin\Phi \nonumber \\
			&& + \epsilon \left[ \delta V - \text{sign}^{m_0}(m_0) J_{m_0}\left(\frac{A}{\Omega}\right) \sin ( k m_z ) \right]
			\Bigg\} ,
			\label{dot_Phi_0}
		\end{eqnarray}
	\end{subequations}
	%
	%	whereas
	%
	\begin{subequations} \label{C2_dot_xi_zeta_0}
		\begin{eqnarray}
			\dot\xi_0 &=& - \alpha \sin\Theta \dot\zeta_0 ,
			%			\frac{ - \alpha \sin\Theta \epsilon \Omega_F }{  1+ \alpha^2+\alpha \delta \epsilon k \Omega_F \sin^2\Theta}
			%			\Bigg(
			%			{\sum_{m=-\infty}^{\infty}}^{m\ne m_0} \text{sign}^m(m) J_{|m|}\left(\frac{A}{\Omega}\right)
			%			\sin \Big[ (V + m \Omega) t - k m_z \Big]
			%			\nonumber \\
			%			%
			%			&& + \delta A \Big[ \cos(\Omega t) - \beta_c \Omega \sin(\Omega t) \Big] \Bigg) ,
			\label{C2_dot_xi_0}\\
			%%%%%%%%%%%%%%%%%%%%%%%%
			\dot\zeta_0 &=&
			\frac{ \epsilon \Omega_F }{  1+ \alpha^2+\alpha \delta \epsilon k \Omega_F \sin^2\Theta}
			\Bigg(
			{\sum_{m=-\infty}^{\infty}}^{m\ne m_0} \text{sign}^m(m) J_{|m|}\left(\frac{A}{\Omega}\right)
			\sin \Big[ (V + m \Omega) t - k m_z \Big]
			+ \delta A \cos(\Omega t)  \Bigg) .
			\label{C2_dot_zeta_0}
		\end{eqnarray}
	\end{subequations}

	Since in this case we have significant contributions of the frequencies $\Omega$ and $V$ to the slow motion velocity $(\dot{\Theta}, \dot{\Phi})$ already in the zeroth order, we do not go to higher orders.
	%In this case we do not go to higher order corrections, since we have a contribution from the Josephson oscillations and external radiation already to the zeroth order slow motion $(\dot\Theta, \dot\Phi)$.
	%From Eqs.~(\ref{dot_Theta_Phi_0}) we obtain
	As in \textit{Case~0}, the equations $\dot\Theta = \dot\Phi = 0$ imply $\Phi = \pi/2$ or $3\pi/2$ (Eqs.~\ref{A0_eq_cond} and \ref{Eq_cond_A0_Phi}), which give an equation for $\Theta$:
	\begin{equation} \label{Eq_cond_C2_Theta}
		\cos\Theta = \epsilon \delta V - \epsilon \text{sign}^{m_0}(m_0) J_{m_0}\left(\frac{A}{\Omega}\right) \sin ( k m_z )
	\end{equation}

\end{widetext}

%\bibliography{references}
%\bibliographystyle{nature}

\end{document}